\documentclass[12pt,a4paper]{article}
\usepackage{eurosym}
\usepackage{setspace}
\usepackage{amssymb}
\usepackage{mathptmx}
\usepackage{hyperref}
\usepackage{graphicx}
\usepackage{amsmath}
\usepackage{amsfonts}
\usepackage[right]{lineno}

\providecommand{\U}[1]{\protect\rule{.1in}{.1in}}
\begin{document}

\title{A Fundamental Duality in the Mathematical and Natural Sciences: From Logic to Biology}
\author{David Ellerman \\
Independent Researcher\\
Ljubljana, Slovenia}
\maketitle

\begin{abstract}
\noindent This is an essay in what might be called ``mathematical metaphysics.'' There is a fundamental duality that run through mathematics and
the natural sciences. The duality starts as the logical level; it is represented by the Boolean
logic of subsets and the logic of partitions since subsets and partitions
are category-theoretic dual concepts. In more basic terms, it starts with the duality between the elements (Its) of  subsets and the distinctions (Dits, i.e., ordered pairs of elements in different blocks) of a partition. Mathematically, the Its $\&$ Dits duality is fully
developed in category theory as the reverse-the-arrows duality. The quantitative versions
of subsets and partitions are developed as probability theory and
information theory (based on logical entropy). Classical physics was based
on a view of reality as definite all the way down. In
contrast, quantum physics embodies (objective) indefiniteness. And finally, there are the two fundamental dual mechanisms at work in biology, the selectionist mechanism and the
generative mechanism, two mechanisms that embody the fundamental duality.

Keywords: Subset-partition duality; logics of subsets and partitions;
category-theory duality; logical entropy; objective
indefiniteness; selectionist and generative mechanisms.
\end{abstract}

\tableofcontents

\section{Introduction: A Fundamental Duality in the Sciences}

There is a fundamental duality that runs through the sciences such as logic,
mathematics (particularly category theory), probability and information
theory, physics, and the life sciences. Historically only one side of
duality has really been developed so the new results are on the development of the little-noticed dual side.

In logic, the highly developed side is based on the Boolean logic of subsets
(often presented in the special case of propositional\ logic). The duality
is well-developed in category theory where the dual to the concept of a
subset, subobject, or `part' is the notion of a quotient set, a quotient
object, or a partition (or, equivalently, an equivalence relation). ``The
dual notion (obtained by reversing the arrows) of `part' is the notion of
partition.'' \cite{law:sfm} (p. 85) so ordinary Boolean algebra is ``the
Algebra of Parts'' \cite{law:sfm} (p. 193). Hence the most basic appearance
of the other side of the duality at the logical level is the new logic of
partitions (\cite{ell:lop}, \cite{ell:intropartitions}, \cite{ell:lop2apps}).

The duality between subsets and partitions can also be expressed, in a more
elementary or granular form, as the duality between the \textit{elements} or
`\textit{Its}' of a subset and the \textit{distinctions} or `\textit{Dits}'
of a partition--where a \textit{distinction} of a partition is an ordered
pairs of elements from the underlying set that are in different blocks of
the partition (or different equivalence classes of the equivalence relation).

On the elements- or Its-side of the duality, the relevant dichotomy is
existence versus nonexistence, e.g., an element is either in a subset or
in the complementary subset. On the distinctions- or Dits-side of the duality, the relevant
dichotomy is distinctions versus indistinctions or, in cognate terms,
inequivalences versus equivalences or distinguishability versus
indistinguishability.

The paper presents the subsets \& partitions duality as it runs through the
mathematical and natural sciences.

\begin{itemize}
\item The most basic form of the duality is in logic, the two logics of the dual notions of
subsets and partitions.

\item Category theory highlights the dual sub-object/quotient-object
architecture that runs throughout mathematics so we develop the basic ideas
in the category of $Sets$. In more general terms, category theory develops
the duality as the ``reverse the arrows'' duality. The only new result is showing how origin of the reverse-the-arrows duality arises in the category of $Sets$ by the interchange of ``elements'' and ``distinctions'' in the definition of a morphism in $Sets$.

\item The next step is the quantitative versions of subsets and partitions
which are probability theory in the case of subsets and logical information
theory (using the notion of logical entropy) in the case of partitions. The formula for logical entropy goes back to the early twentieth century (Corrado Gini) but the development of logical information theory as the quantitative version of partitions is relatively new (\cite{ell:nf4it}, \cite{ell:4open}; \cite{manfredi:4open}).

\item Then we turn to classical physics juxtaposed to quantum mechanics (QM)
where the thesis is that the mathematics (not the physics) of QM is the
Hilbert space version of the mathematics of partitions. That is a new approach to understanding the conceptual origin of the \textit{distinctive} math of QM, i.e., states as vectors in a vector space over $\mathbb{C}$ (which implies the superposition principle) and observables as linear operators on the space \cite{ell:piqr-book}.

\item Finally we extend the duality to the life sciences where it takes the
form of the duality between a selectionist mechanism and a generative
mechanism. The under-developed notion here is the notion of a generative
mechanism where the operative notion of making distinctions is implementing
a code or symmetry-breaking \cite{ell:gen-mech}. The examples of generative mechanisms are not new; what is new is showing how that type of mechanism is the dual of the well-known selectionist mechanism.
\end{itemize}

In short, it was the new developments on the partitions side of the duality that brought the overall duality into view. That duality is the topic of this paper where those new developments on the partition side can only be sketched.

\section{Methods: The Dual Logics of Subsets and Partitions}

While the dual notions of subsets and partitions (or equivalence relations)
are equally fundamental mathematically, the historical development of the
two notions has been very uneven.

\begin{quotation}
\noindent Equivalence relations are so ubiquitous in everyday life that we
often forget about their proactive existence. Much is still unknown about
equivalence relations. Were this situation remedied, the theory of
equivalence relations could initiate a chain reaction generating new
insights and discoveries in many fields dependent upon it.\cite%
{britz:eq-rels} (p. 445)
\end{quotation}

For instance, the notions of join and meet for partitions was known in the
nineteenth century (Dedekind and Schr\"{o}der), but the notion of
implication for partitions was only defined in the twenty-first century \cite%
{ell:lop}. That is, no new operations on partitions were defined throughout
the twentieth century. As noted in 2001, ``the only operations on the family
of equivalence relations fully studied, understood and deployed are the
binary join $\vee$ and meet $\wedge$ operations'' \cite{britz:eq-rels} (p.
445). Incidentally, it might be noted that much of the historical literature 
\cite{birkhoff:lattice-theory3rd} about the ``lattice of partitions'' is
really about the opposite lattice of equivalence relations where the partial
order is inclusion between equivalence relations which is the ``reverse
refinement'' \cite{kung:rotaway} (p. 30) relation between partitions, so the
join and meet are interchanged. In any case, part of the retarded
development of the mathematics of partitions may be due to the notion of a
partition is more complex than the dual notion of a subset. But it may also
be due to the Boolean logic of subsets being almost universally treated in
only the special case of the logic of propositions. Since propositions have
no dual, the whole idea of a dual logic of partitions was not ``in the air.''

We will work with a finite universe set $U=\{ u_{1},...,u_{n}\} $%
, more for convenience than generality. There is a partial order on the set
of all subsets, the \textit{powerset} $\wp( U) $, which is just
the \textit{inclusion} of elements of the subsets. That is, for $S,T\in
\wp( U) $, $S\subseteq T$ if all the elements of $S$ are elements
of $T$. Note that when $S\subseteq T$, then there is a \textit{canonical
injective set function} $S\rightarrowtail T$. The \textit{join} or least
upper bound of subsets $S$ and $T$ is their union $S\cup T$. The \textit{meet%
} or greatest lower bound of subsets $S$ and $T$ is their intersection $%
S\cap T$. The lattice of subsets $\wp( U) $ is the set of all the
subsets with join and meet operations. The lattice also has a \textit{top}
or maximal subset of all elements $U$ and a \textit{bottom} or minimal
subset of no elements $\emptyset$ (the empty set). There is also a
conditional or implication operation on subset $S\Rightarrow T$ (or $%
S\supset T$) which is such that: $S\Rightarrow T=U$ iff (if and only if) $%
S\subseteq T$, i.e., the implication equals the top iff the partial order
holds between the two lattice elements. The subset $S\Rightarrow T=S^{c}\cup
T$ has that property (where $S^{C}=U-S$ is the complement of $S$ in $U$).
The Boolean lattice structure of the joins and meets enriched by the subset
implication or conditional operation makes $\wp( U) $ into a
Boolean algebra.

A \textit{partition} $\pi$ on $U$ is a set of non-empty blocks $\pi=\{
B_{1},...,B_{m}\} $ such that the blocks are disjoint and their union
is all of $U$. The corresponding equivalence relation is $\operatorname*{indit}%
( \pi) =\cup_{j=1}^{m}B_{j}\times B_{j}\subseteq U\times U$ is
the set of ordered pairs of elements that are in the same block of the
partition which are called the \textit{indistinctions} of $\pi$. A \textit{%
distinction} of $\pi$ is an ordered pair of elements in different blocks and
the set of all distinctions is $\operatorname*{dit} ( \pi) =U\times U-%
\operatorname*{indit}( \pi) $. The set of all partitions on $U$ is
denoted $\Pi( U) $ and the \textit{partial order} on it is
defined by refinement, i.e., for another partition $\sigma=\{
C_{1},...,C_{m^{\prime}}\} $, the partition $\sigma$ is \textit{refined%
} by $\pi$, written $\sigma\precsim\pi$, if for every block $B_{j}\in\pi$,
there is a block $C_{j^{\prime}}\in\sigma$ such that $B_{j}\subseteq
C_{j^{\prime}}$. Note that when $\sigma\precsim\pi$, then there is a \textit{%
canonical surjective set function} $\pi\twoheadrightarrow \sigma$ taking
each block $B_{j}\in\pi$ to the block $C_{j^{\prime}}$ that it is contained
in. In terms of distinctions, refinement is equivalent to inclusion of
ditsets, i.e., $\sigma\precsim\pi$ iff $\operatorname*{dit}( \sigma)
\subseteq\operatorname*{dit}( \pi) $.

In the refinement partial order, the \textit{join} $\pi\vee\sigma$ is the
partition whose blocks are all the nonempty intersections $B_{j}\cap
C_{j^{\prime}}$ for $j=1,...,m$ and $j^{\prime}=1,...,m^{\prime}$. The
ditset of the join is just the union of the ditsets, i.e., $\operatorname*{dit}%
(\pi\vee\sigma)=\operatorname*{dit}(\pi) \cup\operatorname*{dit}(\sigma) $. To form the 
\textit{meet} $\pi\wedge\sigma$, take the intersection of all equivalence
relations $E\subseteq U\times U$ such that $\operatorname*{indit}( \pi)
,\operatorname*{indit}( \sigma) \subseteq E$. The intersection of
equivalence relations is always an equivalence relation, and the meet $%
\pi\wedge\sigma$ is the partition whose blocks are the equivalence classes
of the intersection of those equivalence relations. The ditset of the meet $%
\pi\cap\sigma$ is the largest ditset contained in the ditsets of $\pi$ and $%
\sigma$. The join and meet operations turn $\Pi( U) $ into the
lattice of partitions on $U$--which was known in the nineteenth century
(e.g., Richard Dedekind and Ernst Schr\"{o}der). The lattice of partitions
has a top which is the \textit{discrete partition} $\mathbf{1}_{U}=\{
\{ u_{1}\} ,...,\{ u_{n}\} \} $ where all the
blocks are singletons. The bottom is the \textit{indiscrete partition} $%
\mathbf{0}_{U}=\{ U\} $ with only one block $U$. There is an
implication $\sigma\Rightarrow\pi$ which is such that: $\sigma\Rightarrow\pi=%
\mathbf{1}_{U}$ iff $\sigma\precsim\pi$. The partition $\sigma\Rightarrow\pi$
which has that property is like $\pi$ except that for any $B_{j}\in\pi$, if
there is a $C_{j^{\prime}}\in\sigma$ such that $B_{j}\subseteq C_{j^{\prime}}
$, then the block $B_{j}$ is discretized, i.e., replaced by singletons of
all the elements of $B_{j}$. Thus $\sigma \Rightarrow\pi$ is an indicator or
characteristic function for refinement in the sense that if there is a $%
C_{j^{\prime}}$ such that $B_{j}\subseteq C_{j^{\prime}}$, then $B_{j}$ is
replaced by its discrete version $\mathbf{1}_{B_{j}}$, and otherwise $B_{j}$
remains in its indiscrete version $\mathbf{0}_{B_{j}}$. That is why it
satisfies the property: $\sigma \Rightarrow\pi=\mathbf{1}_{U}$ iff $%
\sigma\precsim\pi$. The partition lattice structure of joins and meets
enriched with the partition implication operation makes $\Pi( U) $
in an algebra of partitions.

The Boolean algebra of subsets and the algebra of partitions have been
developed in a way to emphasize the underlying duality of elements of a
subset and distinctions of a partition, i.e., its and dits. The canonical
injections and surjections defined just by the dual logical partial orders
are the ``ur-morphisms'' that define the `canonical' morphisms in the
universal constructions in the category of $Sets$. Table 1 summarizes that
parallelism of the duality.

\begin{center}
\begin{tabular}{l||l|l|}
\cline{2-3}
Its \& Dits & Algebra of subsets $\wp ( U) $ & Algebra of
partitions $\Pi ( U) $  \\ \hline\hline
\multicolumn{1}{|l||}{Its or Dits} & Elements of subsets & Distinctions of
partitions \\ \hline
\multicolumn{1}{|l||}{Partial order} & Inclusion of subsets $S\subseteq T$ & 
Inclusion of ditsets $\operatorname*{dit}( \sigma ) \subseteq \operatorname*{%
dit}( \pi ) $ \\ \hline
\multicolumn{1}{|c||}{Can. maps} & \multicolumn{1}{||c|}{Injection $%
S\rightarrowtail T$} & \multicolumn{1}{|c|}{Surjection $\pi
\twoheadrightarrow \sigma $} \\ \hline
\multicolumn{1}{|l||}{Join} & Union of subsets & Union of ditsets \\ \hline
\multicolumn{1}{|l||}{Meet} & Subset of common elements & Ditset of common
dits \\ \hline
\multicolumn{1}{|l||}{Top} & Subset $U$ with all elements & Partition $%
\mathbf{1}_{U}$ with all distinctions \\ \hline
\multicolumn{1}{|l||}{Bottom} & Subset $\emptyset $ with no elements & 
Partition $\mathbf{0}_{U}$ with no distinctions \\ \hline
Implication & $S\Rightarrow T=U$ iff $S\subseteq T$ & $\sigma \Rightarrow
\pi =\mathbf{1}_{U}$ iff $\sigma \precsim \pi $ \\ \hline
\end{tabular}

Table 1: Elements-and-distinctions (Its \& Dits) duality between the two
logical algebras
\end{center}


\section{Results}

\subsection{Th Fundamental Duality as the Reverse-the-Arrows in Category
Theory}

\subsubsection{The Elements-and-Distinctions Definition of Functions}

Category theory is the foundational theory that brings out the structure or
architectonic of mathematics. Hence we will develop the Its \& Dits duality
in the most basic `ur-category,' the category of $Sets$ (and
functions)--which also underlies the other concrete categories of structured
sets, e.g., groups, rings, modules, vector spaces, and so forth. Since the
morphisms in $Sets$ are set functions, we begin with the natural
elements-and-distinctions definition of set functions.

Given two sets $X$ and $Y$, consider a binary relation $R\subseteq X\times Y$%
.

The relation $R$ is said to \textit{transmit (or preserve) elements} if for all $x\in X$,
there is an ordered pair $( x,y) \in R$ for some $y\in Y$.

The relation $R$ is said to \textit{reflect elements} if for all $y\in Y$,
there is an ordered pair $( x,y) \in R$ for some $x\in X$.

The relation $R$ is said to \textit{transmit (or preserve) distinctions} if for any $%
( x,y) \in R$ and $( x^{\prime},y^{\prime}) \in R$, if 
$x\neq x^{\prime}$, then $y\neq y^{\prime}$.

The relation $R$ is said to \textit{reflect distinctions} if for any $(
x,y) \in R$ and $( x^{\prime},y^{\prime}) \in R$, if $y\neq
y^{\prime}$, then $x\neq x^{\prime}$.

Ordinarily, we might say that a binary relation $R\subseteq
X\times Y$ is the graph of a set function if it is defined everywhere on $X$
and is single-valued in $Y$. But being defined everywhere on $X$ is the same
as transmitting elements and being single-valued in $Y$ is the same as
reflecting distinctions. That gives the elements-and-distinctions definition of a function. 

A \textit{function} is a binary relation that transmits elements and reflects
distinctions. 

The notions of ``transmits'' and ``reflects' give the directionality of the function. The two special types of set functions are injective functions
and surjective functions. They are the functions that satisfy one of the two
other conditions. That is, an \textit{injective} function is one that transmits
distinctions and a \textit{surjective} function is one that reflects elements. In
this manner, we see how the elements-and-distinctions duality provides the
natural concepts to define functions in general and injections and
surjections in particular.

\subsubsection{Subsets and Partitions as Morphisms}

The category theorist, F. William Lawvere, pointed out that every set
function $f:X\rightarrow Y$ determines a subset of the codomain $Y$, namely
its image $f( X) \subseteq Y$ and every set function $%
f:X\rightarrow Y$ also determined a partition on its domain, namely $%
f^{-1}=\{ f^{-1}( y) \} _{y\in f( U) }$.
But unless the function was injective, $f$ would contain extra information
such as the different elements of $X$ that got mapped to a $y\in f(
U) $, and unless a function was surjective, $f$ would contain extra
information such as the elements of $Y$ that had no inverse image, i.e., the
empty fibers $f^{-1}( y) =\emptyset$. Hence in terms of set
functions, a subset was given by an injection and a partition by a
surjection \cite{law:sfm} (p. 86). Furthermore in his introductory text,
Lawvere analyzed the fundamental duality in everyday terms: ``The point of view about maps indicated by
the terms `naming,' `listing,' `exemplifying,' and `parameterizing' is to be
considered as `opposite' to the point of view indicated by the words
`sorting,' `stacking,' `fibering,' and `partitioning'.'' \cite
{law:conceptual} (p. 83)

\subsection{The Canonical Morphisms in Universal Mapping Properties in $Sets$%
}

Category theory isolates the important structures, the universal mapping
properties (UMPs) in mathematics which appear in a dual form, e.g., products
and coproducts as well as equalizers and coequalizers. The dual to a concept is often indicated by the ``co'' prefix. We conjecture that a map is ``canonical'' if relative to the given data, it is reduced to a map defined by the injections or surjections in the two dual logics.  Thus, the
canonical maps in those UMPs are reduced by the given data to the logical injections or
surjections respectively in the algebras of the dual logics of subsets and
partitions. This will be illustrated for coproducts and products (in general see \cite{ell:canonical}).

\subsubsection{Coproduct in $Sets$}

Given two sets $X$ and $Y$ in $Sets$, the idea of the \textit{coproduct} is
to create the set with the maximum number of elements starting with $X$ and $
Y$. Since $X$ and $Y$ may overlap, we must make two copies of the elements
in the intersection. Hence the relevant operation is not the union of sets $
X\cup Y$ but the disjoint union $X\sqcup Y$. To take the disjoint union of a
set $X$ with itself, a copy $X^{\ast}=\{ x^{\ast}:x\in X\} $ of $X
$ is made so that $X\sqcup X$ can be constructed as $X\cup X^{\ast}$. In a
similar manner, if $X$ and $Y$ overlap, then $X\sqcup Y=X\cup Y^{\ast}$.
Then the inclusions $X,Y\subseteq X\sqcup Y$, give the canonical injections $%
i_{X}:X\rightarrow X\sqcup Y$ and $i_{Y}:Y\rightarrow X\sqcup Y$.

The universal mapping property for the coproduct in $Sets$ is that given any
`cocone' of maps $f:X\rightarrow Z$ and $g:Y\rightarrow Z$, there is a
unique map $f\sqcup g:X\sqcup Y\rightarrow Z$ such that $X\overset{i_{X}}{%
\rightarrow}X\sqcup Y\overset{f\sqcup g}{\rightarrow}Z=X\overset{f}{%
\rightarrow}Z$ and $Y\overset{i_{Y}}{\rightarrow}X\sqcup Y\overset{f\sqcup g}%
{\rightarrow}Z=Y\overset{g}{\rightarrow}Z$.

\begin{center}
$%
\begin{array}{ccccc}
X & \overset{i_{X}}{\longrightarrow} & X\sqcup Y & \overset{i_{Y} }{%
\longleftarrow} & Y \\ 
& \searrow^{f} & ^{\exists!}\downarrow^{f\sqcup g} & ^{g}\swarrow &  \\ 
&  & Z &  & 
\end{array}
$

Coproduct diagram
\end{center}

From the data $f:X\rightarrow Z$ and $g:Y\rightarrow Z$, we need to
construct the unique factor map $X\sqcup Y\rightarrow Z$. The map $f$
contributes the image $f( X) $ subset of $Z$ and $g$ contributes
image $g(Y) $ subset of $Z$ so we have the union $f(X) \cup g(Y) \subseteq Z$. To define the canonical factor map 
$f\sqcup g:X\sqcup Y\rightarrow Z$, any $w\in X\sqcup Y$ is either in $%
i_{X}( X) $ so $i_{X}( x) =w$ or is in $i_{Y}(
Y) $ so $i_{Y}( y) =w$ and then $w$ maps by $f\sqcup g$ to $%
f( x) \in f( X) $ or to $g( y) \in g(
Y) $, and $f( X) \cup g( Y) \subseteq Z$. Hence,
in either case, we have the map $w\longmapsto z$ via the injection $f(
X) \cup g( Y) \rightarrow Z$ that makes the triangles
commutes. Both the canonical injections and the factor map were defined by
inclusions, namely $X,Y\subseteq X\sqcup Y$ and $f( X) \cup
g( Y) \subseteq Z$ (the join in the Boolean lattice of subsets on 
$Z$). Thus the canonical maps were defined by the inclusions in $\wp(
X\sqcup Y) $ and $\wp( Z) $.

\subsubsection{Product in $Sets$}

Given two (non-empty) sets $X$ and $Y$ in $Sets$, the product in $Sets$ is
usually constructed as the maximum set of ordered pairs (possible distinctions) of elements from $X$ and $Y$, i.e., the Cartesian product $X\times Y$.

The set $X$ defines a partition $\pi_{X}$ on $X\times Y$ whose blocks are $%
B_{x}=\{ (x,y) :y\in Y\} =\{x\} \times Y$
for each $x\in X$, and $Y$ defines a partition $\pi_{Y}$ whose blocks are $%
B_{y}=\{ ( x,y) :x\in X\} =X\times\{y\}$ for each $%
y\in Y$. Since $\pi_{X},\pi_{Y}\precsim\mathbf{1}_{X\times Y}$, the induced
surjections are the canonical projections $p_{X}:X\times Y\rightarrow X$ and 
$p_{Y}:X\times Y\rightarrow Y$.

The universal mapping property for the product in $Sets$ is that given any
`cone' of maps $f:Z\rightarrow X$ and $g:Z\rightarrow Y$, there is a unique
map $\left\langle f,g\right\rangle :Z\rightarrow X\times Y$ such that $Z%
\overset{\left\langle f,g\right\rangle }{\rightarrow}X\times Y\overset{p_{X}}%
{\rightarrow}X=Z\overset{f}{\rightarrow}X$ and $Z\overset{\left\langle
f,g\right\rangle }{\rightarrow}X\times Y\overset{p_{Y}}{\rightarrow }Y=Z%
\overset{g}{\rightarrow}Y$.

\begin{center}
$%
\begin{array}{ccccc}
&  & Z &  &  \\ 
& \swarrow^{f} & ^{\exists!}\downarrow^{\left\langle f,g\right\rangle } & 
^{g}\searrow &  \\ 
X & \overset{p_{X}}{\longleftarrow} & X\times Y & \overset{p_{Y} }{%
\longrightarrow} & Y%
\end{array}
$

Product diagram
\end{center}

From the data $f:Z\rightarrow X$ and $g:Z\rightarrow Y$, we need to
construct the unique factor map $Z\rightarrow X\times Y$. The map $f$
contributes the inverse-image or coimage $f^{-1}=\{ f^{-1}(
x) :x\in f( Z) \} $ partition on $Z$ and $g$
contributes the coimage $g^{-1}=\{ g^{-1}( y) :y\in g(
Z) \} $ partition on $Z$ so we have the partition join $%
f^{-1}\vee g^{-1}$ whose blocks have the form $f^{-1}( x) \cap
g^{-1}( y) $. To define the unique factor map $\left\langle
f,g\right\rangle :Z\rightarrow X\times Y$, the discrete partition $\mathbf{1}%
_{Z}$ refines $f^{-1}\vee g^{-1}$ so for each singleton $\{ z\} $%
, there is a block of the form $f^{-1}( x) \cap g^{-1}(
y) $ and thus the factor map $\left\langle f,g\right\rangle $ takes $%
z\longmapsto ( x,y) $ and the triangles commute. Both the
canonical projections and the factor map were defined by partition
refinements, namely $B_{X},B_{Y}\precsim\mathbf{1}_{X\times Y}$ and $%
f^{-1}\vee g^{-1}\precsim\mathbf{1}_{Z}$ (the join in the lattice of
partitions on $Z$). Thus the canonical maps were defined by the surjections
in $\Pi( X\times Y) $ and $\Pi( Z) $. 

A similar analysis of the maps that are canonical (using the given information) being provided by the maps from the two partial orders of the dual lattices can be carried out for equalizers and coequalizers and thus for all limits and colimits \cite{ell:canonical} in $Sets$.

\subsubsection{The Duality in $Sets$}

The most abstract form of the fundamental duality is the reverse-the-arrows of category theoretic duality. Given a category like $Sets$ or any category $C$, the reversed arrows in the opposite category $Sets^{op}$ or $C^{op}$, are treated formally or abstractly. But in concrete category of $Sets$ (or any category of structured sets), there are concrete binary relations that serve as the morphisms in the opposite category. One standard example of duality is in plane projective geometry where any proof involving points and lines yields another proof with the points and lines interchanged. Similarly, an arrow-theoretic proof in category theory yields a proof in the opposite category with reversed arrows (morphisms). But in the category of sets, what is interchanged to get the concrete morphisms that serve as the reversed arrows? It is the dual notions of elements (or Its) and distinctions (Dits) that are interchanged to give the dual of a function.

A \textit{cofunction} is a binary relation that transmits distinctions and reflects
elements.

It is easily seen from the structure of the definitions of functions and cofunctions that interchanging elements and distinctions has the same effect as interchanging ``transmits'' and ``reflects,'' which thus reverses the directionality of the morphism or arrow. Cofunctions and functions are different binary relations; they overlap only in the case of isomorphisms. The reverse-the-arrows duality in category theory starts with this interchange of elements and distinctions in $Sets$ to give the concrete category of sets and cofunctions $Sets^{op}$ and then it is abstracted as simply reversed-arrows in abstract categories. In the category of sets and cofunctions, the Cartesian product of sets satisfies the arrow-theoretic definition of the coproduct and the disjoint union of sets satisfies the definition of the product.

In general, category theory thus develops and uses the fundamental duality in abstract
arrow-theoretic (``reverse the arrows'') terms. But it all started in the
`ur-category' of $Sets$ where the arrows are set functions naturally defined
in terms of elements and distinctions--and the dual cofunctions are defined by interchanging the role of elements and distinctions.

\subsection{Probability and Information: The Quantitative Versions of the
Dual Logics}

\subsubsection{Probability Theory}

The next step in the development of the fundamental duality is to develop the quantitative versions of the dual notions of subsets and partitions. The quantitative measure of a subset $S\subseteq U$ is its number of
elements $\vert S\vert $. Probability theory starts with the
assumptions of equiprobability of the elements in $U$, and then the
probability of one draw from $U$ getting an element of event $S$ is the
normalized cardinality of the set:

\begin{center}
$\Pr( S) =\frac{\left\vert S\right\vert }{\left\vert U\right\vert 
}$.
\end{center}

\noindent If the elements have the (always positive) point probabilities $p=(
p_{1},...,p_{n}) $, then $\Pr( S) =\sum_{u_{i}\in S}p_{i}$.

Since probability theory is already well-developed, we turn to information
theory based on the analogous quantitative version of partitions.

\subsubsection{Logical Entropy}

Information theory based on Shannon entropy can define that entropy in terms
of the block probabilities of partitions, e.g., the inverse-image partitions
of random variables \cite{campbell:measure}. But information theorists do
not seem to have exploited the fundamental subset-partition duality.
Gian-Carlo Rota made that key connection. As Gian-Carlo Rota and colleagues
put it: ``The lattice of partitions plays for information the role that the
Boolean algebra of subsets plays for size or probability'' \cite%
{kung:rotaway} (p. 30). In his writings and lectures at MIT, Rota postulated
that:

\begin{center}
$\frac{\text{Probability}}{\text{Subsets}}\approx\frac{\text{Information} }{%
\text{Partitions}}$.
\end{center}

\noindent In his Fubini Lectures, he wrote that since ``Probability is a
measure on the Boolean algebra of events [subsets]'' that gives
quantitatively the ``intuitive idea of the size of a set'', we may ask by
``analogy'' for some measure ``which will capture some property that will
turn out to be for [partitions] what size is to a set.'' He went on to ask:
``How shall we be led to such a property? We have already an inkling of what
it should be: it should be a measure of information provided by a random
variable. Is there a candidate for the measure of the amount of
information?'' \cite{rota:fubini} (p. 67) In view of the subset-partition
duality in terms of elements and distinctions, we know the
``candidate for the measure of the amount of information'' in a partition $
\pi$, namely the number of distinctions $\vert \operatorname*{dit}(
\pi) \vert $--as spelled out in Table 1. Again under the assumption of equiprobable points, the measure of information in $\pi$ is the (normalized) cardinality of its ditset, its \textit{logical entropy}:

\begin{center}
$h( \pi) =\frac{\left\vert \operatorname*{dit}( \pi)
\right\vert }{\left\vert U\times U\right\vert }=\frac{\left\vert U\times U-%
\operatorname*{indit}( \pi) \right\vert }{|U\times U} =1-\sum_{j=1}^{m}%
\frac{\left\vert B_{j}\times B_{j}\right\vert }{\left\vert U\times
U\right\vert }=1-\sum_{j=1}^{m}( \frac{\left\vert B_{j} \right\vert }{%
\left\vert U\right\vert })^{2} =1-\sum_{j=1}^{m}\Pr( B_{j})^{2}
$
\end{center}

\noindent where $\Pr ( B_{j}) =\frac{\left\vert B_{j}\right\vert 
}{\left\vert U\right\vert }$. Now $1=( \sum_{j=1}^{m}\Pr (
B_{j}) ) ^{2}=\sum_{j=1}^{m}\Pr ( B_{j})
^{2}+\sum_{j\neq k}\Pr ( B_{j}) \Pr ( B_{k}) $ for a
general probability distribution $p=( p_{1},...,,p_{n}) $ so $\Pr
( B_{j}) =\sum_{u_{i}\in B_{j}}p_{i}$, and the logical entropy in
the general case is:

\begin{center}
$h( \pi) =1-\sum_{j=1}^{m}\Pr(B_{j})^{2} =\sum_{j\neq
k}\Pr( B_{j}) \Pr( B_{k}) $
\end{center}

\noindent (where it might be noted that each pair of distinct indices is counted twice, e.g., as $\Pr( B_{1}) \Pr( B_{2})$ and as $\Pr( B_{2}) \Pr( B_{1})$). In terms of measure theory, $p$ is a finite probability measure on $U$, so $%
p\times p$ is the product measure on $U\times U$ and then logical entropy is
the value of that measure on the ditset:

\begin{center}
$h( \pi) =p\times p( \operatorname*{dit}( \pi) ) $
\end{center}

\noindent which is the distinctions version of the elements-formula $%
\Pr( S) =p( S) $ for the probability measure $p$ on $U$%
. $Pr(S)$ is the one-draw probability of getting an element of $S$ and $%
h(\pi)$ is the two-draw probability of getting a distinction of $\pi$. Thus the founding of information theory on logical entropy \cite{ell:nf4it} brings out the parallelism between probability theory and information theory provided by the fundamental duality and anticipated by Gian-Carlo Rota.

Since logical entropy is a measure in the sense of measure theory, the
compound notions are defined by the value of that measure on the appropriate
set in $U\times U$:

\begin{itemize}
\item Joint logical entropy of $\pi$ and $\sigma$: $h( \pi\vee
\sigma) =p\times p( \operatorname*{dit}( \pi\vee \sigma)
) =p\times p( \operatorname*{dit}( \pi) \cup\operatorname*{dit}%
( \sigma) ) $;

\item `Conditional' or Difference logical entropy of $\sigma$ minus $\pi$: $%
h( \sigma|\pi) =p\times p( \operatorname*{dit}( \sigma)
-\operatorname*{dit}( \pi) ) $; and

\item Mutual logical entropy of $\pi$ and $\sigma$: $m( \pi
,\sigma) =p\times p( \operatorname*{dit}( \pi) \cap\operatorname*{%
dit}( \sigma) ) $.
\end{itemize}

\noindent Venn diagrams arise from measures, e.g., typically counting
measures. These relationships for logical entropy can be illustrated in the
usual Venn diagram in Figure 1.

\begin{figure}[h]
\centering
\includegraphics[width=0.5\linewidth]{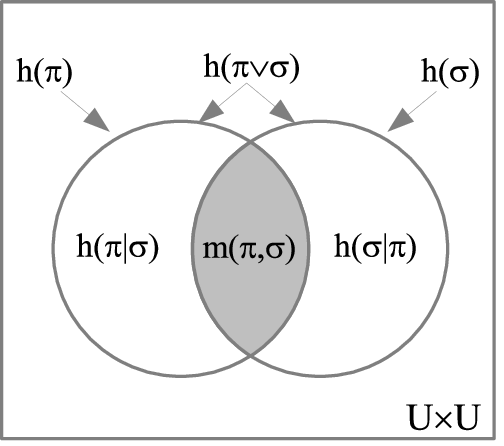} 
\caption{Venn diagram for compound logical entropies}
\label{fig:fig1venn-logical-entropy}
\end{figure}

\subsubsection{The Relationship to Shannon Entropy}

The question immediately arises of what is the relationship of logical
entropy and the well-known \textit{Shannon entropy} which for a partition $%
\pi$ (e.g., the inverse-image partition of a random variable) is:

\begin{center}
$H( \pi) =\sum_{j=1}^{m}\Pr( B_{j}) \log_{2}( 
\frac{1}{\Pr( B_{j}) }) $.
\end{center}

\noindent Any outcome with probability one carries no information (or `surprise') so information is carried by the complements to one. The additive complement to one of $p_{i}$ (i.e., the number added to $p_{i}$ to get $1$) is $1-p_{i}$ and the multiplicative complement to one (i.e., the number multiplied by $p_{i}$ to get $1$) is $\frac{1}{p_{i}}$. The additive probabilistic average of the additive 1-complements is the logical entropy $\Sigma_{i}p_{i}(1-p_{i})$. The multiplicative probabilistic average of the multiplicative 1-complements is the log-free version of the Shannon entropy $\Pi_{i}(\frac{1}{p_{i}})^{p_{i}} $. The appropriate log is then taken to get the additive version of the Shannon entropy, e.g., logs to the base $2$ for coding theory or natural logs for statistical mechanics.

The Shannon entropy is a quantification of information but not a
measure in the sense of measure theory since it is not defined as a measure
on a set. Yet Shannon defined the compound notions of Shannon entropy so
that they satisfied the analogous Venn diagrams. This mystery \cite
{campbell:measure} is explained by the non-linear but monotonic \textit{
dit-to-bit transform} of all the compound logical entropy formulas into the
corresponding formulas for Shannon entropy:

\begin{center}
$1-\Pr( B_{j}) \rightsquigarrow\log_{2}( \frac{1} {\Pr(
B_{j}) }) $
\end{center}

\noindent so that

\begin{center}
$h( \pi) =\sum_{j=1`m}\Pr( B_{j}) ( 1-\Pr(
B_{j}) ) \rightsquigarrow H( \pi)
=\sum_{j=1}^{m}\Pr( B_{j}) \log_{2}( \frac{1}{\Pr(
B_{j}) }) $.
\end{center}

\noindent Since the dit-to-bit transform preserves the Venn diagram
relationships, $h( \pi\vee\sigma) =h( \pi) +h(
\sigma) -m( \pi,\sigma) $ is transformed into the
corresponding relation for the Shannon entropies.

The notion of logical entropy turns up in many fields (see \cite{ell:nf4it}) including bioinformatics or genetic analysis (\cite{nei:diversity}; \cite[chapter 4]{weir:genetics}). For instance,
the sample data may be in the form of the number $N_{ij}=\#$ of ordered
$(  i,j)  $ pairs in the sample. Then the sample statistic for
\textit{heterogeneity} is:

\begin{center}
	$h^{\prime}=\sum_{i}\sum_{j\neq i}\frac{N_{ij}}{N}$.
\end{center}

\noindent If it were $N$ independent draws of ordered pairs from the
probability distribution $p$, then the probability of each pair is $E(
N_{ij})  =p_{i}p_{j}$ so the expected value of the statistic is:

\begin{center}
	$E(  h^{\prime})  =\sum_{i}\sum_{j\neq i}E(  \frac{N_{ij}}%
	{N})  =\sum_{i}\sum_{j\neq i}p_{i}p_{j}=h(  p)  $.
\end{center}

Since probability and logical entropy arise as the quantitative versions of
the dual notions of subsets and partitions, the notion of logical entropy
gives the fundamental or \textit{logical} notion of
information-as-distinctions and the Shannon entropy arises as the transform
that has powerful applications in what Shannon called the ``A Mathematical
Theory of Communication.'' \cite{shannon:comm} The full argument why the
notion of logical entropy provides a \textit{logical} foundation for what is
usually called ``Information Theory'' and provides the definition of
information-as-distinctions has been spelled out elsewhere. (\cite{ell:nf4it}
; \cite{manfredi:4open})

The important point for our purposes at hand is that probability theory and logical
information theory (based on logical entropy) both start with the
quantitative versions of the duality between subsets and partitions--based
on the counting of the elements and distinctions (or Its \& Dits).

\subsection{The Dual Creation Stories: Ex Nihilo and Big Bang}

By moving from bottom up to the top of the dual lattices of subsets and
partitions, we can formulate two very schematic stories of creation. The
stories can be told in terms of the two old metaphysical categories of
substance (or matter) and form \cite{ainworth:form-matter}. Substance and
form are combined in any reality but there are two different ways that the
combination can take place and that yields the two creation stories
illustrated with the lattices of subsets and partitions on a three element
set $U=\{a,b,c\}$ in Figure 2.

\begin{figure}[h]
\centering
\includegraphics[width=0.8\linewidth]{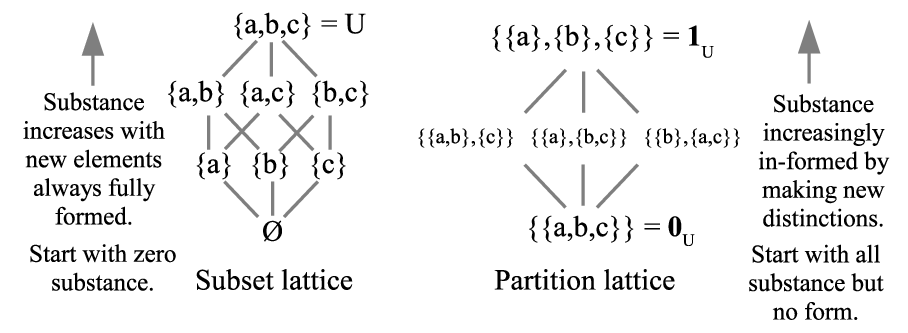}
\caption{Two creation stories told by two dual lattices}
\end{figure}

On the left side of Figure 2 is the story told by moving from bottom to top
in the subset lattice. In the beginning, there was no substance (empty set $%
\emptyset$). The substance was created ex nihilo (new elements) to
eventually reach the universe $U$. Each new element was created in a fully
distinct form so the creation was only in terms of the fully-formed
elements, the new ``its'', going from non-existence to existence. In general, for an element, the question is ``existence or not in a subset.''

On the right side of Figure 2 is the story told by moving from bottom to top
in the partition lattice. In the beginning was all the substance (e.g.,
energy) but with no form (the indiscrete partition $\mathbf{0}_{U}$).

\begin{quotation}
\noindent Just as the Greeks had hoped, so we have now found there is only
one fundamental substance of which all reality consists. If we have to give
this substance a name, we can only call it ``energy.'' But this fundamental
``energy'' is capable of existence in different forms. \cite%
{heisenberg:nuclear}~(p. 116)
\end{quotation}

\noindent That initial state could be described as a state of ``perfect
symmetry.'' \cite{pagels:time} Then the substance was in-formed by the
making of distinctions, i.e., by symmetry-breaking. Thus in this Big-Bang type of creation,
the creation took place by the always-existing substance taking on
information-as-distinctions, the new ``dits'', until the universe was
reached of fully distinct states of the substance (symbolized by the
discrete partition $\mathbf{1}_{U}$). For an ordered pair of elements, the question is ``distinction or not in a partition.''

In the subset creation story, it is the new existence of more ``its'' or
fully-formed elements to eventually reach the full universe of $U$. In the
partition creation story, it is the addition of more ``dits'' or
symmetry-breaking distinctions until the initially unformed substance
eventually reaches the fully-distinct states of $\mathbf{1}_{U}$.

\subsection{Classical Metaphysics}

The classical metaphysics of the always fully definite or fully formed
elements in the subset story of the Boolean lattice of subsets was described by Leibniz in his Principle of
Identity of Indistinguishables (PII) \cite{leib-clarke:letters} (Fourth
letter, p. 22) and by Kant in his Principle of Complete Determination (%
\textit{omnimoda determinatio}).

\begin{quotation}
\noindent Every thing, however, as to its possibility, further stands under
the principle of thoroughgoing determination; according to which, among all
possible predicates of things, insofar as they are compared with their
opposites, one must apply to it \cite{kant:cpr} (B600).
\end{quotation}

\noindent In other words, reality was assumed to be definite `all the way
down,' so if two entities were distinct, then by digging down deep enough,
there would have to be some predicate (i.e., some subset) that would apply to one but not the
other entity. Otherwise, if they were not distinguishable, then there would
not be two entities but one and the same entity as specified in Leibniz's
PII. That principle may fail to hold in the dual partition story. In the
discrete partition $\mathbf{1}_{U}$, the $a$ and $b$ are distinguished in
the separate blocks $\{ a\} $ and $\{ b\} $, but in
the superposition state $\{ a,b\} $, they are not distinguished.
Thus partition logic reproduces Leibniz's principle for the discrete
partition as the ``classical'' part of the partition lattice.:

\begin{center}
For any $u,u^{\prime}\in U$, if $( u,u^{\prime}) \in \operatorname*{%
indit}( \mathbf{1}_{U}) $, then $u=u^{\prime}$.

Partition logic Principle of Identity of Indistinguishables.
\end{center}

\noindent Any other partition in $\Pi( U) $ has non-singleton
blocks in it so the PII does not apply to it. The partition logic PII is
true since no $u\in U$ can be distinguished from itself so the indit set $%
\operatorname*{indit}( \mathbf{1}_{U}) $ of the discrete partition
consists only of the the self-pairs $( u_{i},u_{i}) $ for $%
i=1,...,n$.

The subset creation story may correspond to some older notions of \textit{ex nihilo}
creation, but the theory of creation in modern physics is the Big Bang which
clearly corresponds to the partition story. The characteristic feature of classical physics and of our intuitive view of the macroscopic world is that it is fully definite. In the philosophy of physics discussions, the full-definiteness is sometime known as full ``haecceity'' (\cite{teller:haecceity}; \cite{stachel:hole}). But on the other side of the duality, there is indefiniteness or ``quiddity'' without full haecceity, e.g., in quantum mechanics.

\subsection{Quantum Mechanics Math as the Hilbert Space Version of Partition
Math}

\subsubsection{Introduction: A Logical Basis for Superposition}

Quantum mechanics (QM) has a distinctive type of mathematics, i.e., all vectors are states (which implies the superposition principle) and observables are operators, quite different
from the math of classical mechanics. Our analysis is of that distinctive
math of QM, not the physics. The thesis is that the math of QM is the
Hilbert space version of the math of partitions, or, put the other way
around, partition math is a bare-bones, schematic, or skeletal version of QM
math. The notion of a superposition state is the basic notion in QM that
separates it from the fully-definite or definite-all-the-way-down
metaphysics of classical mechanics. When referring to a quantum particle
(not the classical notion of a particle) as a ``quanton,'' Mario Bunge makes
that point.

\begin{quotation}
\noindent Another surprising peculiarity of quantons is that they are blurry
or fuzzy rather than neat or sharp. Whereas in classical physics all
properties are sharp, in quantum physics only a few are: most are blunt or
smudged. ... The reason for this fuzziness is that ordinarily an isolated
quanton is in a ``coherent'' state, that is, the combination or
superposition (weighted sum) of two or more basic states (or
eigenfunctions). The superposition or ``entanglement'' of states is a
hallmark of quantum mechanics \cite{bunge:m-and-m} (pp. 49-50).
\end{quotation}

\noindent If quantum field theory is also included, then James Cushing makes
the same point, namely that ``superposition, with the attendant riddles of
entanglement and reduction, remains the central and generic interpretative
problem of quantum theory'' \cite{cushing:qft} (p. 34).

Our thesis about QM math provides the \textit{logical} basis to interpret
superposition in terms of indefiniteness since partitions provide the
logical model of the indefiniteness of the states in a non-singleton block
of a partition, i.e., in a non-singleton equivalence class in an equivalence
relation. Given a superposition state $\alpha\vert a \rangle+\beta\vert b\rangle$, the \textit{support} (forget the vector space machinery) is the set $\{a,b\}$, so the schematic set-version of a superposition state is its support (as a non-singleton equivalence class or block in a partition). This thesis has been argued at length in papers (\cite{ell:ftm}, 
\cite{ell:new-theories}) and a book \cite{ell:piqr-book}. Hence we will only
summarize some of the salient points here.

\subsubsection{Quantum States}

We will demonstrate the thesis by briefly describing the partition math
version of quantum states, quantum observables, and quantum state reduction
(`measurement'). The mathematical tool that brings out the partitional
aspects of quantum states is not the state vector representation but the
density matrix representation. Hence we construct the density matrix version
of a partition $\pi=\{B_{1},...,B_{m}\} $ on a set $U$ with
positive point probabilities $p=( p_{1},...,p_{n}) $. $U$ is interpreted as the set of possible \textit{eigenstates} of a quantum particle (``eigen'' is interpreted as ``definite''). For each
block $B_{j}$, let $\left\vert b_{j}\right\rangle $ be the $n$-ary real
column vector with the $i^{th}$ entry being $\sqrt{\frac{p_{i}}{\Pr(
B_{j}) }}$ if $u_{i}\in B_{j}$ and $0$ otherwise. These vectors are
normalized and, since the blocks are disjoint, the vectors are orthogonal to
each other so $\left\langle b_{k}|b_{j}\right\rangle =\delta_{jk}$ (the
Kronecker delta where $\delta_{jk}=1$ if $j=k$ and $0$ otherwise). Then the $%
n\times n$ density matrix $\rho( B_{j}) $ is constructed as the
outer product of $\left\vert b_{j}\right\rangle $ with its transpose $%
\left\vert b_{j}\right\rangle ^{t}=\left\langle b_{j}\right\vert $:

\begin{center}
$\rho( B_{j}) =\left\vert b_{j}\right\rangle \left\langle
b_{j}\right\vert $.
\end{center}

\noindent The entries in $\rho( B_{j}) $ are $\rho(
B_{j}) _{ik}=\frac{\sqrt{p_{i}p_{k}}}{\Pr( B_{j}) }$ if $%
u_{i},u_{k}\in B_{j}$, else $0$. Then the density matrix $\rho(
\pi) $ for the partition is the probabilistic sum of the density
matrices for the blocks:

\begin{center}
$\rho( \pi) =\sum_{j=1}^{m}\Pr( B_{j}) \rho(
B_{j}) $.
\end{center}

\noindent The entries in $\rho( \pi) $ are $\rho( \pi)
_{ik}=\sqrt{p_{i}p_{k}}$ if $( u_{i},u_{k}) \in\operatorname*{indit}%
( \pi) $, else $0$, so the non-zero entries of $\rho(
\pi) $ correspond to the ordered pairs in the equivalence relation $%
\operatorname*{indit}( \pi) $ and the zeros correspond to the ordered
pairs in $\operatorname*{dit}( \pi) $. If $\rho( \pi) _{ik}>0
$ ( $i\neq k$), then the $u_{i}$ and $u_{k}$ are blurred or cohered
together in one `superposition' block. Those non-zero off-diagonal elements,
indicating the presence of superposition in the corresponding diagonal
elements, are called ``coherences'' in QM and they allow the characteristic
interference effects.

\begin{quotation}
\noindent For this reason, the off-diagonal terms of a density matrix ...
are often called ``quantum coherences'' because they are responsible for the
interference effects typical of quantum mechanics that are absent in
classical dynamics \cite{auletta:qm} (p. 177).
\end{quotation}

\noindent A density matrix $\rho$ represents a \textit{pure} state if $%
\rho^{2}=\rho$, otherwise a \textit{mixed} state. All the $\rho(
B_{j}) $ are pure states and the only partition with a pure state
density matrix is $\mathbf{0}_{U}$. Any density matrix is positive Hermitian
so its $n$ eigenvalues are non-negative reals and sum to one. In the case of 
$\rho( \pi) $, the eigenvalues are the $m$ block probabilities $%
\Pr( B_{j}) $ and $n-m$ zeros. In the case of a pure density
matrix such as $\rho( \mathbf{0}_{U}) $ or $\rho(
B_{j}) $, there is one eigenvalue of $1$ with the rest of the
eigenvalues of zero. Given any $\rho( B_{j}) $ [$\rho( 
\mathbf{0}_{U}) $ being the special case where $B_{1}=U$], the vector $
\vert b_{j}\rangle $ is recovered (up to sign) as the normalized eigenvector
associated with the eigenvalue of $1$, and $\rho( B_{j})
=\vert b_{j}\rangle \langle b_{j}\vert $ follows as
the spectral decomposition of the density matrix.

Taking $S=B_{j}$, a pure state density matrix $\rho( S) $ for a
subset $S\subseteq U$ has the normalized eigenvector $\vert
s\rangle $ associated with the eigenvalue of $1$. The probability of
drawing $u_{i}$ given $S$ is given by the formula: $\Pr( u_{i}|S)
=\langle u_{i}|s\rangle ^{2}$--which shows the origin of the Born
Rule at the set level. Hence that vector $\vert s\rangle $ plays
the role of the state vector or (non-wavy) `wave function.' at the set level.

These properties of partition math formulated using the density matrices $
\rho(\pi) $ of partitions all hold in the Hilbert space math of
QM. Those corresponding properties are summarized in Table 2.

\begin{center}
\begin{tabular}{|c|c|}
\hline
Partition math & Quantum math \\ \hline\hline
Density matrix: $\rho(\pi)$ & $\rho$ \\ \hline
ON vectors: $\langle b_{j^{\prime}}|b_{j}\rangle=\delta_{jj^{\prime}}$ & $%
\langle u_{i^{\prime}}|u_{i}\rangle=\delta_{ii^{\prime}}$ \\ \hline
Eigenvalues: $\Pr(B_{1}),...,\Pr(B_{m}),0,...,0$ & $\lambda_{1},...,\lambda
_{n}$ \\ \hline
Spectral decomp.: $\rho(\pi)=\sum_{j=1}^{m}\Pr(B_{j})|b_{j}\rangle\langle
b_{j}|$ & $\rho=\sum_{i=1}^{n}\lambda_{i}|u_{i}\rangle\langle u_{i}|$ \\ 
\hline
Non-zero off-diag. entry: Cohering of diag. states & Cohering of diag. states
\\ \hline
Pure state: $\rho( S) =\left\vert s\right\rangle \left\langle
s\right\vert $ & $\rho( \psi) =\left\vert \psi\right\rangle
\left\langle \psi\right\vert $ \\ \hline
Eigenvector Eigenvalue 1 State vector: $\left\vert s\right\rangle $ & $%
\left\vert \psi\right\rangle $ \\ \hline
Born Rule: $\Pr( u_{i}|S) =\left\langle u_{i}|s\right\rangle ^{2}$
& $\Pr( u_{i}|\psi) =\left\vert \left\langle u_{i}
|\psi\right\rangle \right\vert ^{2}$ \\ \hline
\end{tabular}

Table 2: Quantum states: Partition math and QM math
\end{center}

\subsubsection{Quantum Observables}

There is (in the mathematical folklore) a semi-algorithmic procedure to
associate vector space concepts with the corresponding set concepts. For
instance, a subspace is the vector space concept that corresponds to the set
concept of a subset. We call this procedure, the:

\begin{center}
\textit{Yoga of Linearization}.\ 

Given a basis set $U$ of a vector space,

consider it first as just a set, apply a set concept to the set $U$,

and then take the vector space notion linearly generated by it

as the corresponding vector space concept.
\end{center}

The Yoga of Linearization can be viewed as an embellishment on the free
vector space functor from the category of $Sets$ to the category of vector
spaces over a given field, i.e., $\mathbb{C}$ for our application to QM. A
subset $S$ generates a subspace $\left[ S\right] $ and the cardinality of
the subset $\left\vert S\right\vert $ corresponds to the dimension $%
\dim( \left[ S\right] ) $ of the subspace. Given a partition $\pi$
on $U$ as a set, each block $B_{j}$ generates a subspace $\left[ B_{j}\right]
$ and the collection $\{ \left[ B_{j}\right] \} _{j=1}^{m}$
constitutes a direct-sum decomposition (DSD) of the vector space where a DSD
of a vector space is a set of subspaces so that each non-zero vector in the
space can be uniquely represented as a sum of (non-zero) vectors from the
subspaces. In particular, those vectors in the sum are the non-zero
projections of the vector to the subspaces.

Thus we may say that the vector space version of a set partition is a DSD.
Moreover, we could have defined a partition $\pi$ on $U$ as a set of subsets 
$\{ B_{j}\} _{j=1}^{m}$ so that each non-empty subset of $U$ can
be uniquely represented as the union of non-empty subsets of the $B_{j}$s.
If the union of the $B_{j}$s was not all of $U$, then the difference $%
U-\cup_{j=1}^{m}B_{j}$ would have no representation as a union of non-empty
subsets of the $B_{j}$s, and if $B_{j}\cap B_{k}\neq\emptyset$, then that
overlap would have two representations.

An observable is a Hermitian (or self-adjoint) operator on a Hilbert space $%
F:V\rightarrow V$ which will have real eigenvalues. The set version is a
numerical attribute $f:U\rightarrow\mathbb{R}$ where $U$ is a basis set for $%
F$. Given any numerical attribute $f:U\rightarrow\mathbb{R}$, a Hermitian
operator $F$ is defined on $V$ by the definition $Fu_{i}=f(
u_{i}) u_{i}$ (or $F\left\vert u_{i}\right\rangle =f(
u_{i}) \left\vert u_{i}\right\rangle $ if we use the Dirac notation)
on the basis $U$ and then linearly extended to the whole space. Or given a
Hermitian operator $F:V\rightarrow V$ and an orthonormal basis $U$ of
eigenvectors of $F$, the numerical attribute is recovered as the eigenvalue
function $f:U\rightarrow\mathbb{R}$ that assigns to each eigenvector its
eigenvalue. The numerical attribute $f:U\rightarrow\mathbb{R}$ has the
inverse-image partition $f^{-1}=\{ f^{-1}( r) \}
_{r\in f( U) }$ and the eigenspaces for the $F$ defined by $f$
are the subspaces $\left[ f^{-1}( r) \right] $ generated by the
blocks $f^{-1}( r) $ for the eigenvalues $r\in f( U) $.

What is the set notion of an eigenvector? For a subset $S\subseteq U$ and
real $r\in\mathbb{R}$, let ``$rS$'' stand for the statement ``the value of $f
$ on the subset $S$ is $r$'', so that ``$f\upharpoonright S=rS$'' ($%
f\upharpoonright S$ is $f$ restricted to $S$) is the set version of the
eigenvalue equation: $F\left\vert u_{i}\right\rangle = r\left\vert
u_{i}\right\rangle $. Thus the set notion of an eigenvector is just a
constant set of a numerical attribute and its eigenvalue is that constant
value on the set. A characteristic or indicator function $%
\chi_{S}:U\rightarrow\{ 0,1\} \subseteq\mathbb{R}$, where $%
\chi_{S}( u_{i}) =1$ if $u_{i}\in S$, else $0$, defines the
projection operator $P_{\left[ S\right] }:V\rightarrow V$ to the subspace
generated by the subset $S$. Thus characteristic functions on sets correlate
with projection operators on vector spaces. Moreover, each observable $F$
with the eigenvalues $\lambda _{1},...,\lambda_{n}$ and eigenspaces $\{
V_{\lambda_{i}}\} $ has a spectral decomposition $F=\sum_{i=1}^{n}%
\lambda_{i}P_{V_{i}}$. Hence the corresponding spectral decomposition of a
numerical attribute $f:U\rightarrow \mathbb{R}$ is $f=\sum_{r\in f(
U) }r\chi_{f^{-1}( r) }:U\rightarrow\mathbb{R}$.

Applied to observables, our thesis that the QM math of observables is the
Hilbert space version of the partition math of numerical attributes over the
reals. Those correlations between the partition math of numerical attributes
and QM math of observables are given in Table 3.

\begin{center}
\begin{tabular}{|c|c|}
\hline
Partition math $f:U\rightarrow\mathbb{R} $ & Hilbert space math $%
F:V\rightarrow V$ \\ \hline\hline
{\small Partition} $\{ f^{-1}( r) \} _{r\in f(
U) }$ & {\small DSD }$\{ V_{r}\} _{r\in f( U) }$
\\ \hline
$U=\sqcup_{r\in f( U) }f^{-1}( r) $ & $V=\oplus _{r\in
f( U) }V_{r}$ \\ \hline
Numerical attribute $f:U\rightarrow\mathbb{R}$ & Observable $Fu_{i}=f(
u_{i}) u_{i}$ \\ \hline
$f\upharpoonright S=rS$ & $Fu_{i}=ru_{i}$ \\ \hline
Constant set $S$\ of $f$ & Eigenvector $u_{i}$\ of $F$ \\ \hline
Value $r$\ on constant set $S$ & Eigenvalue $r$\ of eigenvector $u_{i} $ \\ 
\hline
Characteristic fcn. $\chi_{S}:U\rightarrow\{ 0,1\} $ & Projection
operator $P_{\left[ S\right] }u_{i}=\chi_{S}(u_{i})u_{i}$ \\ \hline
Spec. Decomp. $f=\sum_{r\in f( U) }r\chi_{f^{-1}(
r) }$ & Spectral Decomp. $F=\sum_{r\in f( U) }rP_{V_{r}} $
\\ \hline
Set of $r$-constant sets $\wp( f^{-1}( r) ) $ & 
Eigensp. $V_{r}=\left[ f^{-1}( r) \right] $ of $r$ -eigenvect. \\ 
\hline
\end{tabular}

Table 3: Partition math for $f:U\rightarrow\mathbb{R}$ and corresponding QM
math for $F:V\rightarrow V$.
\end{center}

\subsubsection{Quantum Measurement}

Given an observable $F:V\rightarrow V$ with an ON (Ortho-Normal) basis of eigenvectors $U$,
an eigenvalue function $f:U\rightarrow\mathbb{R}$, a DSD of eigenspaces $%
\{ V_{r}\} $ associated with the eigenvalues $r\in f(
U) $, the projective measurement of a state $\left\vert \psi
\right\rangle $ with density matrix $\rho$ is described by the L\"{u}ders
mixture operation (\cite{luders:meas}; \cite{furry:luders}) which produces a
mixed state density matrix

\begin{center}
$\hat{\rho}=\sum_{r\in f( U) }P_{V_{r}}\rho P_{V_{r}}$.

Hilbert space L\"{u}ders Mixture Operation
\end{center}

\noindent where $P_{V_{r}}$ is the projection to the eigenspace $V_{r}=\left[
f^{-1}( r) \right] $. To see the set version, we start with the
numerical attribute $f:U\rightarrow\mathbb{R}$ where the $n\times n$
projection matrices $P_{r}$ for $r\in f( U) $ are diagonal
matrices with the diagonal entries given by $(P_{r})_{ii}=\chi_{f^{-1}(
r) }(u_{i})$. Then the set version of the L\"{u}ders mixture operation
on a density matrix $\rho( \pi) $ is given by:

\begin{center}
$\hat{\rho}(\pi)=\sum_{r\in f( U) }P_{r}\rho( \pi)
P_{r}$.

Partition version of L\"{u}ders mixture operation
\end{center}

\noindent It is then easily shown \cite{ell:ftm} that $\hat{\rho}(\pi
)=\rho( \pi\vee f^{-1}) $. Thus the set version of the L\"{u}ders
mixture operation is density matrix for the join of two partitions, $\pi$
representing the state being measured, and $f^{-1}=\{ f^{-1}(
r) \} _{r\in f( U) }$ representing the observable.

These results, which only give a small part of the partition math underlying
QM math \cite{ell:piqr-book}, are summarized in Table 4.

\begin{center}
\begin{tabular}{l||l|l|}
\cline{2-3}
Dictionary & Partition math & Hilbert space math \\ \hline\hline
\multicolumn{1}{|l||}{Notion of state} & $\rho( \pi) =\sum
_{j=1}^{m}\Pr( B_{j}) \left\vert b_{j}\right\rangle \left\langle
b_{j}\right\vert $ & $\rho=\sum_{i=1}^{n}\lambda_{i}\left\vert u_{i}
\right\rangle \left\langle u_{i}\right\vert $ \\ \hline
\multicolumn{1}{|l||}{Notion of observable} & $f=\sum_{r\in f( U)
}r\chi_{f^{-1}( r) }:U\rightarrow\mathbb{R} $ & $F=\sum_{r\in
f( U) }rP_{V_{r}}$ \\ \hline
\multicolumn{1}{|l||}{Notion of measurement} & \multicolumn{1}{||c|}{$\hat {%
\rho}( \pi) =\sum_{r\in f( U) }P_{r}\rho(
\pi) P_{r}$} & $\hat{\rho}=\sum_{r\in f( U) }P_{V_{r}}\rho
P_{V_{r}}$ \\ \hline
\end{tabular}

Table 4: Three basic notions: Partition version and QM math version.
\end{center}

\subsubsection{The Objective Indefiniteness Interpretation of QM}

The partition math basis for QM \textit{mathematics} shows a new way to handle the
century-old problem of interpreting the QM formalism. The ``cutting at the joint'' between QM math and QM physics is indicated by the absence of Planck's constant in our analysis that deals only with the math of vector spaces and Hilbert spaces in particular. The all-important superposition principle (that the sum of two quantum states is another possible quantum state) and Dirac's use of CSCOs (Complete Sets of Commuting Observable) do not involve Planck's constant.

Partitions (or
equivalence relations) are the math to model distinctions and indistinctions
and thus to model indefiniteness (states of a particle in a non-singleton
`superposition' block of a partition or equivalence class of an equivalence
relation) as opposed to definite or eigen-states (singleton blocks as in the
discrete partition $\mathbf{1}_{U}$). This approach to understanding QM
corroborates an interpretation by Heisenberg, Shimony, and many others who
see quantum reality (like the part of an iceberg under the water) that is
characterized by objective indefiniteness.

\begin{quotation}
\noindent The conceptual elements of quantum theory that now underlie our
picture of the physical world include objective chance, quantum
interference, and the objective indefiniteness of dynamical quantities.
Quantum interference, which is directly observable, was readily absorbed by
the physics community. Objective chance and indefiniteness, being of more
philosophical significance, gained acceptance only after much debate and
conceptual analysis, when it was recognized that observed phenomena are
better understood through these notions than through older ones or hidden
variables \cite{jaeger:qobjects} (p. vii).
\end{quotation}

Heisenberg, Shimony, Jaeger, and others may describe an indefinite
superposition as being a ``potentiality'' as opposed to an actuality but
that should be interpreted as a manner of speaking about indefiniteness
rather than as a different ontological category. There is only one
ontological category of reality but the real state may be indefinite between
a number of definite or eigen-states.

The Feynman rules \cite{jaeger:qobjects} (pp. 110-111) specify that making
the change from indefinite to more definite is by making distinctions.
Different levels of indefiniteness may be schematically pictured, in an 
\textit{anschaulich} (intuitive) manner, using a lattice of partitions where
a state reduction (or `measurement') moves upward (`vertically') in the
lattice from indefinite to more definite states--which von Neumann called a
Type I quantum process. The Type II quantum process is a unitary
transformation that moves horizontally at the same level of indefiniteness
transforming one basis set $\{ a,b,c\} $ into another basis set $%
\{ a^{\prime },b^{\prime},c^{\prime}\} $ as pictured in Figure 3.
In the schematic terms of the lattice of partitions, Figure 3 shows the
classical part of reality (fully definite states as the ``tip of the
iceberg'') and the quantum reality involving indefinite superposition states
(like the ``underwater part of an iceberg'').

\begin{figure}[h]
\centering
\includegraphics[width=0.9\linewidth]{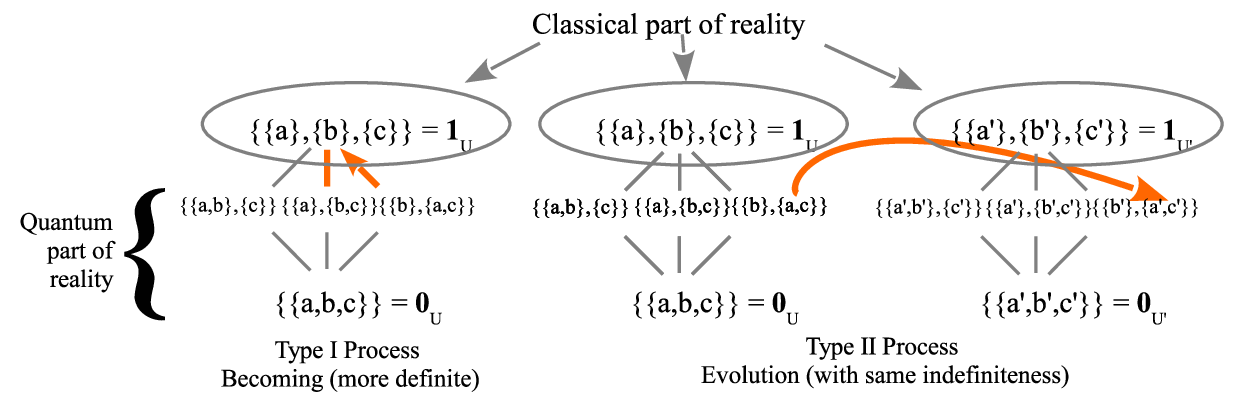} 
\caption{The two von Neumann processes illustrated schematically using
partition lattices}
\label{fig:3}
\end{figure}

The idea of the two basic processes in QM has worried some quantum
philosophers. Classical mechanics has no superposition states, only fully
definite states, and only one type of fundamental process that transforms
the definite states into other definite states.

The schematic picture of Figure 3 shows how it is natural to have two
fundamental processes, the vertical process of going from indefinite to more
definite and the horizontal process of moving at the same level of
indefiniteness. Moreover, this shows why it is natural to have only one type
of fundamental process in classical mechanics. We have seen in the partition
logic Principle of Identity of Indistinguishables that classicality is
represented by the discrete partition $\mathbf{1}_{U}$. But at that
classical level, there can be no more vertical movement from indefinite to
more definite since the classical states are fully definite--so there is
only the horizontal movement from definite states to other definite states.

It should also be noted that boundary between state reductions
(``measurements'') and unitary evolution is specified in the Feynman rules
in terms of distinguishability and indistinguishability--concepts modeled at
the logical level by partitions. The Feynman rules were stated in his work in the early 1950s, e.g., \cite{feynman:1951}.

\subsubsection{Commuting, Non-commuting, and Conjugate Operators}

The non-commuting or even conjugate operators of QM math at first seem to
have little connection with partition math. But each observable operator has
the associated direct-sum decomposition of eigenspaces, and DSDs are the
vector space version of partitions. Suppose we have two observables $%
F,G:V\rightarrow V$ with the respective DSDs of eigenspaces $\{
V_{i}\} _{i\in I}$ and $\{ W_{j}\} _{j\in J}$. We know that
the operation on partitions to create more distinctions is the join so we
consider a join-like operation on the two DSDs to yield the set of non-zero
subspaces $\{ V_{i}\cap W_{j}\} $. Partitions on the same set (or
numerical attributes on the set) are said to be \textit{compatible}, and the
join of two partitions on a set $U$ is always another partition on $U$. But
these subspaces of simultaneous eigenvectors may not span the whole space $V$%
. Let $\mathcal{SE}$ be the space spanned by the simultaneous eigenvectors
in the non-zero spaces $V_{i}\cap W_{j}$. Then it is a theorem (\cite%
{ell:ftm}, \cite{ell:piqr-book}) that $F$ and $G$ commute iff $\mathcal{SE}=V
$, and $F$ and $G$ are conjugate iff $\mathcal{SE}=\{0\} $ (the zero
space), i.e., they have no simultaneous eigenvectors. 

Thus commutativity
depends solely on the vector-space partitions (DSDs), not on the operators \textit{per se}. In vector spaces like $\mathbb{Z}_{2}^{n}$, the only operators are projection operators (with eigenvalues of $0$ or $1$), but DSDs can have up to $n$ subspaces and the DSDs can be commuting, non-commuting, or even conjugate. The
join-like operation on DSDs is only properly called a \textit{join} in the
case of commutativity.

The Heisenberg indeterminacy principle is usually stated in a quantitative form involving Planck's constant, but the underlying fact that there are conjugate DSDs with no simultaneous eigenvectors (i.e., $\mathcal{SE}=\{0\} $) is a fact about vector spaces that has nothing to do with Planck's constant \cite{ell:piqr-book}.

One of the basic operations on partitions that we will see in many contexts
is the join of enough partitions to reach the discrete partition, i.e., to
distinguish all the elements of $U$. If the partitions are the
inverse-images of numerical attributes then we have one of the word-for-word
translation dictionaries between partition math and the QM math (where Planck's constant plays no role).

\textbf{Set case}: A set of compatible numerical attributes $%
f,g,...,h:U\rightarrow\mathbb{R}$ is said to be \textit{complete} (a
Complete Set of Compatible Attributes or CSCA) if the join of their
inverse-image partitions is the partition with all blocks of cardinality
one. Then each element $u_{i}\in U$ is uniquely specified by the ordered set
of its values.

\textbf{QM case}: A set of commuting observables $F,G,...,H:V\rightarrow V$
is said to be \textit{complete} (a Complete Set of Commuting Observables or
CSCO \cite{dirac:principles}) if the join of their eigenspace DSDs is a DSD
with all subspaces of dimension one. Then each simultaneous eigenvector is
uniquely specified by the ordered set of its eigenvalues.

\subsubsection{Group Representation Theory}

There is one mathematical theory, group representation theory, that is
particularly applicable to quantum mechanics and particle physics. That is
because a group is essentially a `dynamic' way to define an equivalence
relation. An equivalence relation on a set is reflexive, symmetric, and
transitive. As Hermann Weyl pointed out: ``The three postulates for a group
simply state that each figure is similar to itself and that similarity is
symmetric and transitive (see the axioms for equivalence on p. 9)'' \cite%
{weyl:pmns} (p. 73).

Given a group $G$ and a set $U$, a \textit{representation} of $G$ on $U$ (or
a $G$ group action on $U$) is a set of isomorphisms (i.e., permutations) $%
\{ R_{g}:U\rightarrow U\} _{g\in G}$, such that (1) for the
identity $e\in G$, $R_{e}$ is the identity map on $U$, (2) for any $g\in G$
with its inverse $g^{-1}$, $R_{g}R_{g^{-1}}=R_{g^{-1}}R_{g}=R_{e}$, and (3)
For $g$, $g^{\prime}$, $g^{\prime\prime}\in G$, if $gg^{\prime}=g^{\prime
\prime}$ then $R_{g}R_{g^{\prime}}=R_{g^{\prime\prime}}$. When $G$ acts on $U
$, then it defines a partition, the \textit{orbit partition}. For any $u\in U
$, the orbit, block, or equivalence class containing $u$ is the set $\{
u^{\prime}\in U:\exists R_{g},R_{g}( u) =u^{\prime}\} $ of
elements of $U$ that can be reached by the action of some $R_{g}$. In other
words, the actions of the group $R_{g}(u)=u^{\prime}$ are creating the
indistinctions $( u,u^{\prime}) $ of the orbit partition. Often a
group is described as a \textit{symmetry} \textit{group} so if $R_{g}(
u) =u^{\prime}$ then $u$ is said to be \textit{symmetric} to $%
u^{\prime}$--so that being symmetric is the equivalence relation whose
equivalence classes are the orbits.

In the development of the math of partitions, we have seen that a
(non-discrete) partition can be refined by adding more distinctions, e.g., $%
\pi$ can be refined to $\pi\vee\sigma$ by adding the new distinctions of $%
\sigma$ since $\operatorname*{dit}( \pi\vee\sigma) =\operatorname*{dit}(
\pi) \cup\operatorname*{dit}( \sigma) $ so the new distinctions
are $\operatorname*{dit}( \sigma) -\operatorname*{dit}( \pi) $. If 
$H$ is a subgroup of $G$, then $\{ R_{g}:U\rightarrow U\} _{g\in
H}$ is a group representation of $H$ on $U$ and since it has no
indistinctions $R_{g}( u) =u^{\prime}$ for $g\in G-H$, its orbit
partition will refine the orbit partition of the $G$-representation. This
way to create more distinctions is called ``\textit{symmetry breaking}''; it
creates smaller and thus less indefinite or more definite orbits. In the
lattice of partitions, the most refined partition is the discrete partition $%
\mathbf{1}_{U}$, and it is the orbit partition of the smallest subgroup $%
\{ e\} $ consisting of just the identity $e$.

As would be expected from the Yoga of Linearization, the set concepts
linearize to vector space representations of a group. Given a group $G$ and
a vector space $V$ over $\mathbb{C}$, a vector space representation of $G$
is a set of invertible linear maps $\{ R_{g}:V\rightarrow V\} $
satisfying $R_{e}=I$ and $R_{g}R_{g^{\prime}}=R_{gg^{\prime}}$. These vector
space representations have very important applications in quantum mechanics 
\cite{weyl:groups-qm} and particle physics \cite{chen:group-reps}--as would
be expected since a group representation is a `dynamic' way to define a
partition (of orbits) in the set case and a direct-sum decomposition (of
irreducible subspaces of $V$) in the vector space case. The approach to
isolating the irreducible representation or irreps (representations
restricted to irreducible subspaces) developed by the Nanjing School of J.
Q. Chen and colleagues is particularly appropriate for our purposes since
``the foundation of the new approach is precisely the theory of the complete
set of commuting operators (CSCO) initiated by Dirac...'' \cite
{chen:group-reps} (p. 2).

It is well beyond the scope of this paper to go into the other major theory of modern physics, general relativity, but suffice it to say that indefiniteness plays a key role there as well as emphasized by the general reletivity theorist, John Stachel. 

\begin{quotation}
	 So both relativity and quantum theory lead to the same conclusion: Leibniz's principle is not universally applicable. There is a category of entities with quiddity but no inherent haecceity. Given that both general relativity and quantum mechanics are based on such entities, it is difficult to believe that, in any theory purporting to underlie both relativity and quantum theory, inherent individuality would re-emerge in its fundamental entities, whatever they are... . \cite{stachel:hole} (p. 55)
\end{quotation}

\subsection{Selectionist and Generative Mechanisms in the Life Sciences}

\subsubsection{Introduction: The Basic Ideas}

The elements-and-distinctions or Its $\&$ Dits duality leads in the life sciences to two
types of mechanisms, the well-known selectionist mechanism and the `dual'
mechanism that will be called the ``generative mechanism.''

The \textit{selectionist mechanism}, abstractly described, is a process that
constantly whittles down sets of \textit{actual} entities or elements, e.g.,
the set of random variations of a type of organism, to subsets that are
selected according to some fitness criterion.

In contrast, a \textit{generative mechanism} operates on some relatively
undifferentiated entity (a root or stem) containing a number of \textit{potential} outcomes
so that making distinctions will generate a variety of different possible
outcomes. The making of distinctions can be conceptualized as the
implementation of a code or as symmetry-breaking.

The question of existence or non-existence on one side of the duality is dual to the question of distinction or indistinction on the other side.

The following Figure 4 abstractly illustrates the different mechanisms:

\begin{itemize}
\item the selectionist mechanism of starting with a set of actual distinct
entities and reducing it by selections (according to some fitness criteria)
to a smaller or even singleton subset, versus

\item the generative mechanism of starting with a relatively
undifferentiated entity (analogous to a superposition state) that embodies
various possibilities or potentialities which then can be generated by
repeated distinctions (or symmetry-breakings) to in-form a more definite
specific outcome.
\end{itemize}

\begin{figure}[h]
\centering
\includegraphics[width=0.7\linewidth]{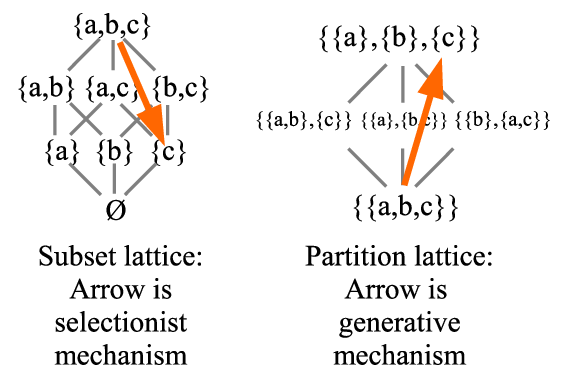} 
\caption{Abstract description of the two dual mechanisms using the two dual
lattices}
\label{fig4}
\end{figure}

Since the selectionist type of mechanism is already well known and much
promoted (\cite{dennett:darwin}; \cite{cziko:womiracles}), we will focus
mostly on developing the relevant concepts to describe generative mechanisms.

\subsubsection{Partitions and Codes}

We live in an `Information Age' so we begin by showing how the machinery of
information coding embodies generative mechanisms. Mathematically, a
partition on a set represents one way to differentiate the elements of the
set into different blocks. The join with another partition generates a
partition with more refined (smaller) blocks that makes all the distinctions
of the partitions in the join. Starting from a single block consisting of
the set of all possibilities like the unbranched root of the tree
(symbolized by the indiscrete partition $\mathbf{0}_{U}$), a sequence of
partitions joined together differentiates all the elements of the set
ultimately into singleton blocks (i.e., $\mathbf{1}_{U}$) that are the
leaves of the tree. All the (instantaneous) codes of coding theory can be
generated in this way and then the codes are implemented in practice to
traverse the tree to generate the coded outcomes (e.g., messages).

With consecutive joins of partitions (always on the same universe set), the
blocks get smaller and smaller until they reach the discrete partition $%
\mathbf{1}_{U}$ (like in a CSCA or CSCO) with the smallest non-empty blocks
being the singletons of elements of $U$. The least refined partition is the
indiscrete partition $\mathbf{0}_{U}=\{ U\} $ whose only block is
all of $U$ and it represents the root (or stem as in stem cell) of the tree.

The tree that would illustrate the consecutive joins in Table 5 where $%
U=\{ a,b,c\} $ consists of three leaves or messages. Since the
code is binary, all the partitions to be joined are binary with the first
block on the left labeled with the code letter $0$ and the other block is
labeled $1$ as in $\{\overset{0}{\{a\}},\overset{1}{\{b,c\}\}}$, i.e., it is
like a numerical attribute on $U$ taking values in the set of code letters
with the code letter assigned to a block being the value of the attribute on
those elements of $U$. When a message first appears in the Consecutive Joins
column as a singleton, then its history of $0$'s and $1$'s in the second
column gives its code.

\begin{center}
\begin{tabular}{|c|c|c|c|}
\hline
& Partitions to be Joined & Consecutive Joins (tree) & Codes \\ 
\hline\hline
$1$ & $\{\{a\},\{b,c\}\}$ & $\{\{a\},\{b,c\}\}$ & $0=$ (code for) $a$ \\ 
\hline
$2$ & $\{\{a,b\},\{c\}\}$ & $\{\{a\},\{b\},\{c\}\}$ & $10=b$, $11=c$ \\ 
\hline
\end{tabular}

Table 5: Instantaneous codes for $U=\{ a,b,c\} $ generated by
consecutive joins
\end{center}

\noindent In Figure 5, the partition joins are indicated and the trajectory
from the complete `superposition' state $\mathbf{0}_{U}$ at the root of the
tree to the messages is given in the (upside down) tree diagram with the
rows of Table 5 indicated.

\begin{figure}[h]
\centering
\includegraphics[width=0.7\linewidth]{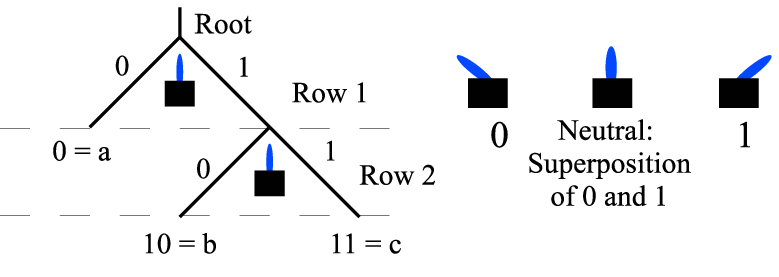} 
\caption{The code tree with switches to represent reduction of indefinite
states to more definite states}
\label{fig5}
\end{figure}

At each junction in the tree, there is pictured a switch which (to borrow
the language from QM) reduces the superposition state to one of the two more
definite outcomes. That is, the first switch at the root $\mathbf{0}%
_{U}=\{ \{ a,b,c\} \} $ reduces it to one of the more
definite states in $\{ \{ a\} ,\{ b,c\} \} $
and then the second switch reduces the superposition $\{ b,c\} $
to $\{ b\} $ or $\{ c\} $. The final result is the
fully definite states of $\mathbf{1}_{U}$, i.e., the leaves in the code tree.

For a more complex example, consider the five messages in $U=\{
u_{1},...,u_{5}\} $. To generate a binary code for the five outcomes
we consider the repeated joins of binary partitions in Table 6. Think of the
block on the left as representing the code letter $0$ and the block on the
right as representing the code letter $1$. In the repeated joins of binary
partitions, the blocks get smaller and smaller until a singleton block is
reached for each message--as we saw before in CSCAs and CSCOs. When a
message first appears as a singleton (i.e., fully differentiated outcome) in
the Consecutive Joins column representing the sequence of more and more
refined partitions, then the sequence of $0$-blocks or $1$-blocks in the
Partitions column containing that specific outcome give the code for that
outcome or message \cite{martin-england:entropy} (p. 56).

\begin{center}
\begin{tabular}{|c|c|c|c|}
\hline
& Partitions to be joined & Consecutive Joins (tree) & Codes \\ 
\hline\hline
$1$ & $\{\{u_{1}\},\{u_{2},u_{3},u_{4},u_{5}\}\}$ & $\{\{u_{1}\},\{u_{2}
,u_{3},u_{4},u_{5}\}\}$ & $0=$ (code for) $u_{1}$ \\ \hline
$2$ & $\{\{u_{1},u_{2},u_{3}\},\{u_{4},u_{5}\}\}$ & $\{\{u_{1}\},\{u_{2}
,u_{3}\},\{u_{4},u_{5}\}\}$ &  \\ \hline
$3$ & $\{\{u_{1},u_{2},u_{3},u_{4}\},\{u_{5}\}\}$ & $\{\{u_{1}\},\{u_{2}
,u_{3}\},\{u_{4}\},\{u_{5}\}\}$ & $110=u_{4},111=u_{5}$ \\ \hline
$4$ & $\{\{u_{1},u_{2},u_{4}\},\{u_{3},u_{5}\}\}$ & $\{\{u_{1}\},\{u_{2}
\},\{u_{3}\},\{u_{4}\},\{u_{5}\}\}$ & $1000=u_{2},1001=u_{3} $ \\ \hline
\end{tabular}

Table 6: Instantaneous codes generated by consecutive partition joins.
\end{center}

For instance, the $u_{1}$ message first appears as a singleton in the first
row where it was in the $0$-block so its code word is just $0$. No
singletons appear in the second join (second row) so there are no two-letter
code words in the developing code. Then in the third join (row $3$) both $%
u_{4}$ and $u_{5}$ first appear as singletons in the Consecutive Joins
column so their history of $0$-blocks and $1$-blocks (starting in row $1$
Partitions column) give their codes of $110=u_{4}$ and $111=u_{5}$. Finally $%
u_{2}$ and $u_{3}$ appear in singletons in the final join (row $4$) where
all outcomes are singletons in $\mathbf{1}_{U}$, and their history of $0$%
-blocks and $1$-blocks gives their codes of $1000=u_{2}$ and $1001=u_{3}$.
The history of each outcome or message to its singleton cannot be repeated
for any other message (since singletons cannot further differentiate) so
this procedure always generates what is called an \textit{instantaneous}
code where no code word can be the prefix of another code word \cite%
{ell:nf4it} (pp. 62-64).

Figure 6 gives the `progress' of an outcome starting with its
undifferentiated form in the root of the `upside down' tree (the indiscrete
partition) and then traced out as each outcome or message code is
implemented to finally yield the fully distinguished outcome, i.e., its
singleton block in the Consecutive Joins column of Table 6.

\begin{figure}[h]
\centering
\includegraphics[width=0.7\linewidth]{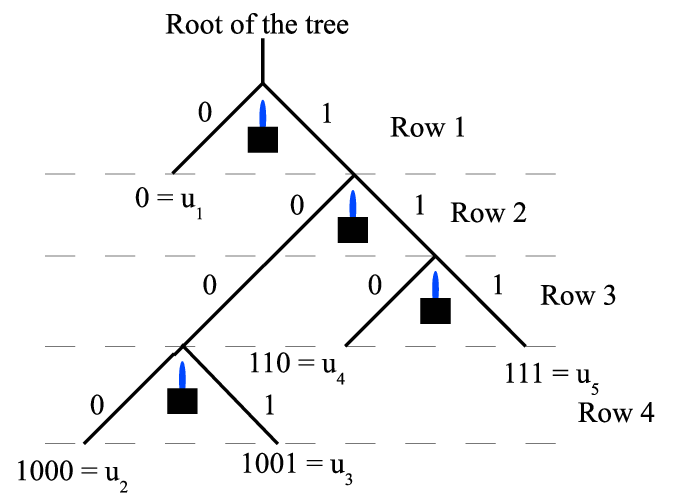} 
\caption{Code tree corresponding to Table 6}
\label{fig6}
\end{figure}

\subsubsection{The genetic code}

The most famous code is, of course, the genetic code which is instantaneous
so it can be generated by a sequence of partition joins. In this case, each
partition has four blocks corresponding to the four code letters U, C, A,
and G in the code alphabet. For the partitions in Figure 7, which correspond
to the partitions in the Partitions column like in Table 6, the consecutive
joins give all $64$ singletons after three branchings or joins so the amino
acids have 3-letter code words. Empirically, the code is redundant since
there can be several codes for the same acid.

The circles in Figure 7 trace out the code for Thr4 (one of the code words
for Thr, Threonine) which is ACG = Thr4. Note that the order of the
partitions counts in the consecutive-joins determination of the genetic
codes. A different ordering gives a different code which may not describe
the operation of the DNA-RNA machinery to produce a certain amino acid from
a given code word.

\begin{figure}[h]
\centering
\includegraphics[width=0.7\linewidth]{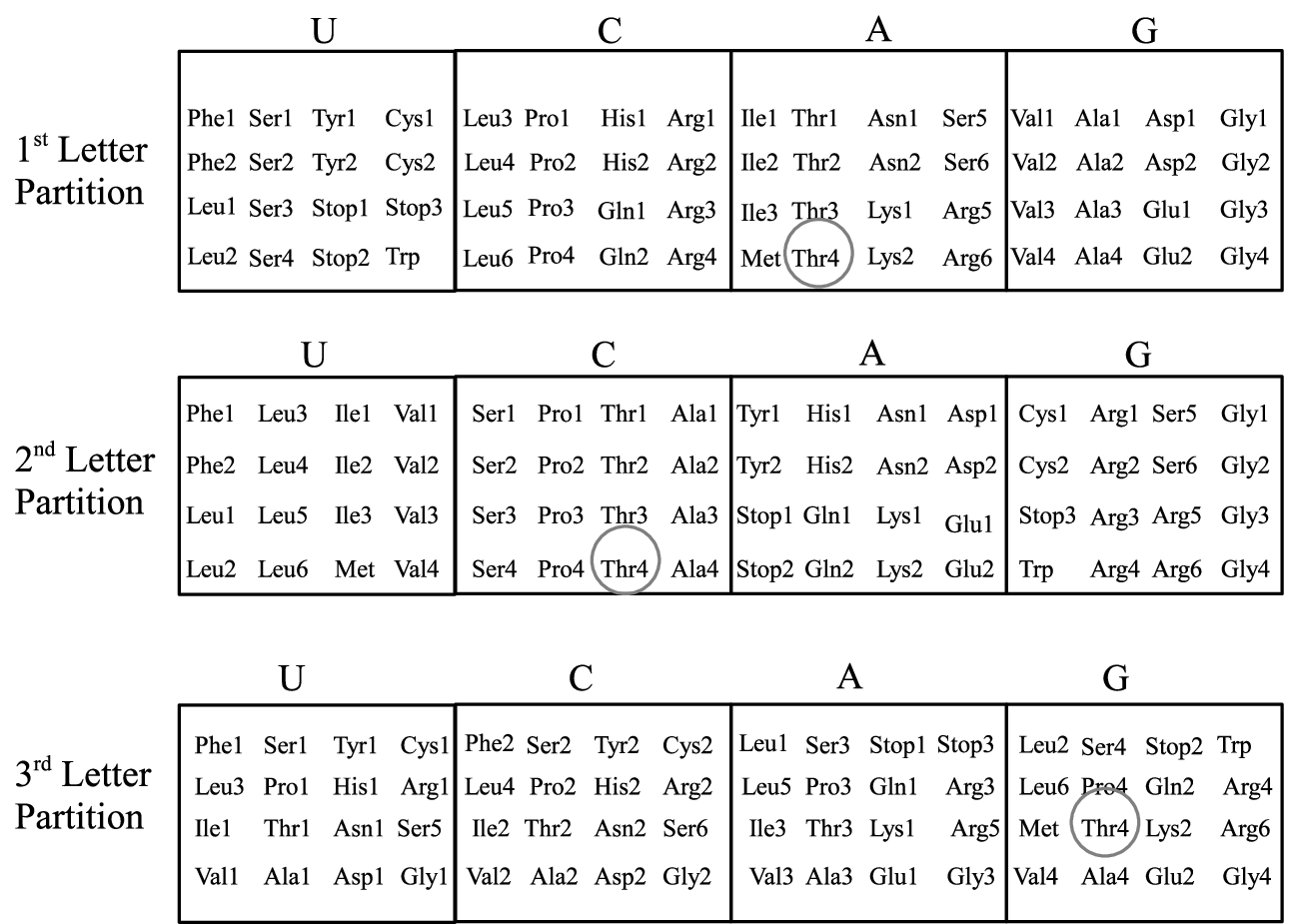} 
\caption{The three partitions that generate the genetic code}
\label{fig7}
\end{figure}

In terms of a tree diagram as in Figure 8, the tree would branch four ways
at each branching point and there are three levels, so there are $4^{3}=64$
leaves in the tree.

\begin{figure}[h]
\centering
\includegraphics[width=0.4\linewidth]{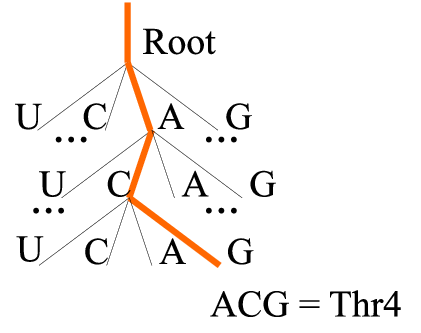} 
\caption{Tree representation (partial) of code implementation for ACG = Thr4}
\label{fig8}
\end{figure}

\noindent The generative mechanism associated with the genetic code is the
whole DNA-RNA machinery that generates the amino acid as the output from the
code word as the input. If we abstractly represent the DNA-RNA machinery as
that tree with $64$ leaves, then the given code word tells the machinery how
to traverse the tree to arrive at the desired leaf.

\subsubsection{The Principles \& Parameters Mechanism for Language
Acquisition}

Noam Chomsky's Principles \& Parameters (P\&P) mechanism (\cite%
{chomp-lasnik:pandp}; \cite{chomsky:minimalist}) for language learning can
be modeled as a generative mechanism. Again, we can consider a tree diagram
where each branching point has a two-way switch to determine one grammatical
rule or another in the language being acquired.

\begin{quotation}
\noindent A simple image may help to convey how such a theory might work.
Imagine that a grammar is selected (apart from the meanings of individual
words) by setting a small number of switches - 20, say - either ``On'' or
``Off.'' Linguistic information available to the child determines how these
switches are to be set. In that case, a huge number of different grammars
(here, 2 to the twentieth power) will be prelinguistically available,
although a small amount of experience may suffice to fix one \cite{higginbotham} (p. 154).
\end{quotation}

And the reference to 20 recalls the game of ``20 questions'' where the
answers to the yes-or-no questions guides one closer and closer to the
desired hidden answer. Chomsky uses the Higginbotham model to describe a
Universal Grammar (UG) as a generative mechanism.

\begin{quotation}
\noindent Many of these principles are associated with parameters that must
be fixed by experience. The parameters must have the property that they can
be fixed by quite simple evidence, because this is what is available to the
child; the value of the head parameter, for example, can be determined from
such sentences as John saw Bill (versus John Bill saw). Once the values of
the parameters are set, the whole system is operative. Borrowing an image
suggested by James Higginbotham, we may think of UG as an intricately
structured system, but one that is only partially ``wired up.'' The system
is associated with a finite set of switches, each of which has a finite
number of positions (perhaps two). Experience is required to set the
switches. When they are set, the system functions \cite{chomsky:knowoflang}
(p. 146).
\end{quotation}

In the tree modeling of the P\&P approach, the relative poverty of
linguistic experience that sets the switches plays the role of the code that
guides the mechanism from the undifferentiated root state (all switches at
neutral) to the final specific grammar represented as a leaf.

\begin{quotation}
\noindent Most important of all, it offered an explanatory model for the
empirical analyses which opened a way to meet the challenge of ``Plato's
Problem'' posed by children's effortless ``yet completely successful''
acquisition of their grammars under the conditions of the poverty of the
stimulus. This becomes particularly clear if we take the view that
parametric variation exhausts the possible morphosyntactic variation among
languages and further assume that there is a finite set of binary
parameters. Imposing an arbitrary order on the parameters, a given
language's set of parameter settings can then be reduced to a series of $0$s
and $1$s, i.e. a binary number $n$ \cite{roberts:ug} (p. 17).
\end{quotation}

\noindent The binary number $n$ is the code to traverse the tree down to the
leaf representing the particular grammar.

The question about the acquisition of a grammar is a good topic to compare
and contrast a selectionist mechanism with a generative mechanism. What
would a selectionist approach to learning a grammar look like? A child would
(perhaps randomly) generate a diverse range of babblings, some of which
would be differentially reinforced or selected by the linguistic environment
(e.g., \cite{skinner:behave}).

\begin{quotation}
\noindent Skinner, for example, was very explicit about it. He pointed out,
and he was right, that the logic of radical behaviorism was about the same
as the logic of a pure form of selectionism that no serious biologist could
pay attention to, but which is [a form of] popular biology -- selection
takes any path. And parts of it get put in behaviorist terms: the right
paths get reinforced and extended, and so on. It's like a sixth grade
version of the theory of evolution. It can't possibly be right. But he was
correct in pointing out that the logic of behaviorism is like that [of na\"{\i}ve adaptationism], as did Quine \cite{chomp-mcgil:interview} (Section
10).
\end{quotation}

A more sophisticated version of a selectionist model for the
language-acquisition faculty or universal grammar (UG) could be called the
format-selection (FS) approach (Chomsky, private communication). The diverse
variants that are actualized in the mental mechanism are different sets of
rules or grammars. Then given some linguistic input from the linguistic
environment, the grammars are evaluated according to some evaluation metric,
and the best rules are selected.

\begin{quotation}
\noindent Universal grammar, in turn, contains a rule system that generates
a set (or a search space) of grammars, $\{G_{1},G_{2},\ldots,G_{n}\}$. These
grammars can be constructed by the language learner as potential candidates
for the grammar that needs to be learned. The learner cannot end up with a
grammar that is not part of this search space. In this sense, UG contains
the possibility to learn all human languages (and many more). ... The
learner has a mechanism to evaluate input sentences and to choose one of the
candidate grammars that are contained in his search space \cite%
{nowak-komarove} (p. 292)
\end{quotation}

After a sufficient stream of linguistic inputs, the mechanism should
converge to the best grammar that matches the linguistic environment. Since
it is optimizing over sets of rules, this model at least takes seriously the
need to account for the choice of rules (rather than just assuming the child
can infer the rules from raw linguistic data). Early work (through the
1970s) on accounting for the language-acquisition faculty or universal
grammar (UG) seems to have assumed such an approach. The problems that
eventually arose with the FS approach could be seen as the conflict between
descriptive and explanatory adequacy.

Since selection operates on actualities, in order to describe the enormous
range of human language grammars, the range of grammars considered would
make for an unfeasible computational load of evaluating the linguistic
experience. If the range was restricted to make computation more feasible,
then it would not explain the variety of human languages. Hence the claim is
that the P\&P generative mechanism gives a more plausible account of human
language acquisition than a behavioral/selectionist approach.

\subsubsection{Embryonic stem cell development}

Our simple partition lattice or rooted tree models of a generative mechanism
pale beside the complexity of embryonic development. Nevertheless, it seems
clear that the stem cells have the role of embodying the potentialities like
the indiscrete partition $\mathbf{0}_{U}$ or the root in a rooted tree.
Thus, the role of stem cells in the development of an embryo from a
fertilized egg into a full organism can be modeled as a generative mechanism.

\begin{figure}[h]
\centering
\includegraphics[width=0.3\linewidth]{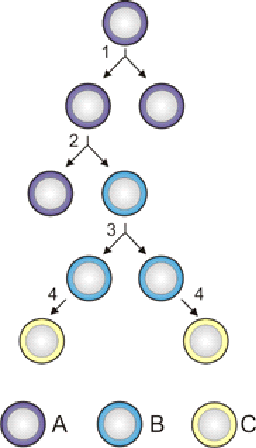} 
\caption{Stem cell division and differentiation [Attribution: Peter
Znamenskiy, CC BY-SA 3.0 https://en.wikipedia.org/ at ``Stem cell'']}
\label{fig9}
\end{figure}

As illustrated in Figure 9, stem cells come in three general varieties: A)
the stem cells that can reproduce undifferentiated copies of themselves, B)
the stem cells that can reproduce but can also produce a somewhat
differentiated cell, and C) a specialized differentiated cell. Each
branching point in a tree has a certain number of possible leaves or
terminal types of cells beneath it in the tree. In a division (\#1) of an
A-type cell, each of the resulting A-type cell could have a full set of
leaves beneath it. But when it splits (\#2) into another A-type cell and a
B-type cell, then the B-cell has a restricted number of leaves beneath it.
The B-type cells can split (\#3) in two, and finally when a B-type cell
gives rise (\#4) to a specific C-type of cell, that is a terminal branch,
i.e., a leaf, in the tree.

The codes that inform the progress through the tree are not fully
understood, but apparently the positional epigenetic information in the
developing embryo provides the information about the next development steps.
In general terms,

\begin{quotation}
\noindent\lbrack t]hat model harks back to the ``developmental landscape''
proposed by Conrad Waddington in 1956. He likened the process of a cell
homing in on its fate to a ball rolling down a series of ever-steepening
valleys and forked paths. Cells had to acquire more and more information to
refine their positional knowledge over time --- as if zeroing in on where
and what they were through ``the 20 questions game, according to Jan\'{e}
Kondev, a physicist at Brandeis University. \cite{cepel:mathcells}
\end{quotation}

Again, the reference to the game of 20 questions reveals the common
generative mechanism of traversing a tree from the root to a specific leaf.
Information is distinctions so more and more distinctions (``forked paths'')
are made along a path like the path in the partition lattice from the one
block in the indiscrete partition to smaller and smaller blocks until
finally arriving at a singleton block in the discrete partition.

In Figure 10, the lattice of partitions on $U=\{a,b,c,d\}$ is represented
using the shorthand of eliminating the innermost curly brackets in favor of
juxtaposition so $\{\{a\},\{b,c,d\}\}$ is $\{a,bcd\}$. The path is indicated
where the block containing the $b$ outcome is differentiated by more and
more distinctions until finally becoming fully distinct as a singleton block
in the discrete partition. The indicated path through the lattice of
partitions is like the Consecutive Joins column in Tables 5 and 6. The
increasing amount of information used to make all the differentiations is
indicated by the rising logical entropies of the increasingly refined
partitions.

\begin{figure}[h]
\centering
\includegraphics[width=0.8\linewidth]{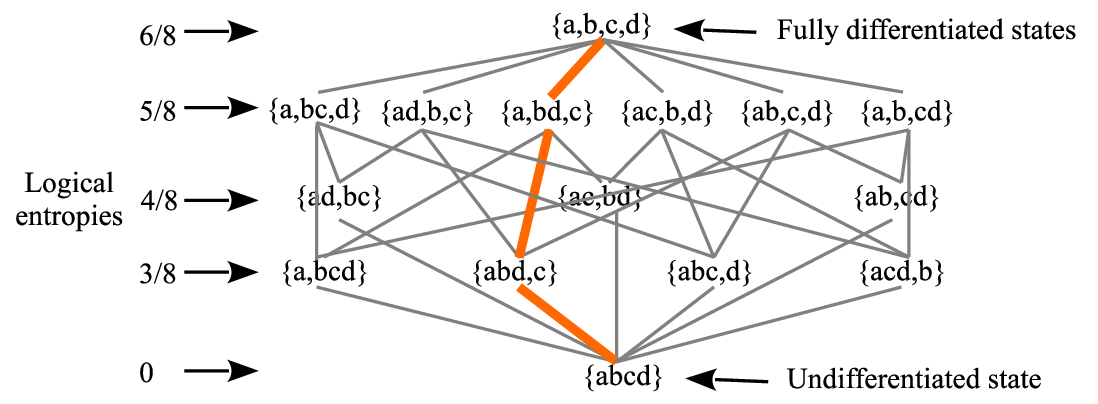} 
\caption{One developmental path of $b$ from the undifferentiated beginning
to fully distinct outcome}
\label{fig10}
\end{figure}

Moreover, we have seen in the analysis of group representations that
symmetries play the role of equivalences or indistinctions, and thus that
the making of distinctions is described as ``symmetry-breaking.'' That holds
true also in embryonic development.

\begin{quotation}
Ultimately, symmetry breaking shapes your whole body, from the location of
your head and toes to the position of your organs, from the symmetric
location of lungs and kidneys to the way the heart is on the left. All this,
in turn, derives from asymmetries on the molecular scale.

Symmetry breaking is essential to shape many of the most dramatic phases of
our development \cite{zernicka:dance} (p. 13).
\end{quotation}

Thus, it seems clear that the whole complex and only partly understood
process of development from a stem cell to an fully differentiated organism
can be described as a generative mechanism.

\subsubsection{Selectionist and Generative Mechanisms Redux}

There is a long tradition in biological thought of juxtaposing selectionism,
associated with Darwin, with instructionism, associated with Lamarck (\cite%
{medawar:reith}; \cite{jerna:antibodies}). In an instructionist or
Lamarckian mechanism, the environment would transmit detailed instructions
about a certain adaptation to an organism, while in a selectionist
mechanism, a diverse variety of (random) variations would occur, and then
some variations would be selected by the environment as the ``survival of
the fittest.'' The discovery that the immune system was a selectionist
mechanism \cite{jerne:nat-sel} generated a wave of enthusiasm, a ``Second
Darwinian Revolution'' \cite{cziko:womiracles}, for selectionist theories 
\cite{dennett:darwin}.

In his Nobel Lecture \cite{jerne:nobel}, Niels Jerne even tried to draw
parallels between Chomsky's generative grammar and selectionism. One of the
distinctive features of a selectionist mechanism is that the possibilities
must be in some sense actualized or realized in order for selection to
operate on and differentially amplify or select some of the actual variants
while the others languish, atrophy, or die off. In the case of the human
immune system, ``It is estimated that even in the absence of antigen
stimulation a human makes at least $10^{15}$ different antibody
molecules---its preimmune antibody repertoire'' \cite{alberts-et-al:mbio}
(p. 1221).

In Chomsky's critique of a selectionist theory of universal grammar, he
noted the computational infeasibility of having representations of all
possible human grammars in order for linguistic experience and an evaluation
criterion to perform a selective function on them. The analysis of Chomsky's
P\&P theory as a \textit{generative} mechanism instead suggests that the old
juxtaposition of ``selectionism versus instructionism'' is \textit{not} the
most useful framing for the study of biological mechanisms. It is better
framed as selectionist mechanisms versus generative mechanisms.

The discovery of the genetic code and DNA-RNA machinery for the production
of amino acids powerfully showed the existence of another biological
mechanism, a generative mechanism, that is quite distinct from a
selectionist mechanism. The examples of Chomsky's P\&P theory of grammar
acquisition and the role of stem cells in embryonic development provide more
evidence of the importance of generative mechanisms.

To better illustrate these two main types of biological mechanisms, it might
be useful to illustrate a selectionist and a generative mechanism in solving
the same problem of determining one among the $8=2^{3}$ options considered
in Figure 10. The eight possible outcomes might be represented as: $%
|000\rangle$ , $|100\rangle$, $|010\rangle$, $|110\rangle$, $|001\rangle$, $%
|101\rangle$, $|011\rangle$, $|111\rangle$.

In the selectionist scheme, all eight variants are in some sense actualized
or realized in the initial state so that a fitness criterion or evaluation
metric (as in the FS scheme) can operate on them. Some variants do better
and some worse as indicated by the type size in Figure 11.

\begin{figure}[h]
\centering
\includegraphics[width=0.7\linewidth]{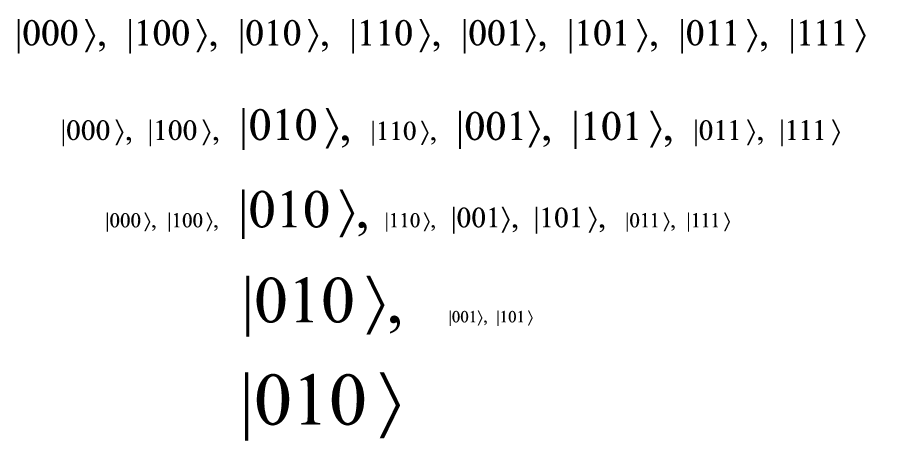} 
\caption{A selectionist determination of the outcome $|010\rangle$}
\label{fig11}
\end{figure}

\noindent The ``unfit'' options dwindle, atrophy, or die off leaving the
most fit option $|010\rangle$ as the final outcome.

With the generative mechanism, the initial state (the root of the tree) is
where all the switches are in neutral, so all the eight potential outcomes
are in a ``superposition'' (between left and right) state indicated by the
plus signs in the following Figure 12.

\begin{figure}[h]
\centering
\includegraphics[width=0.7\linewidth]{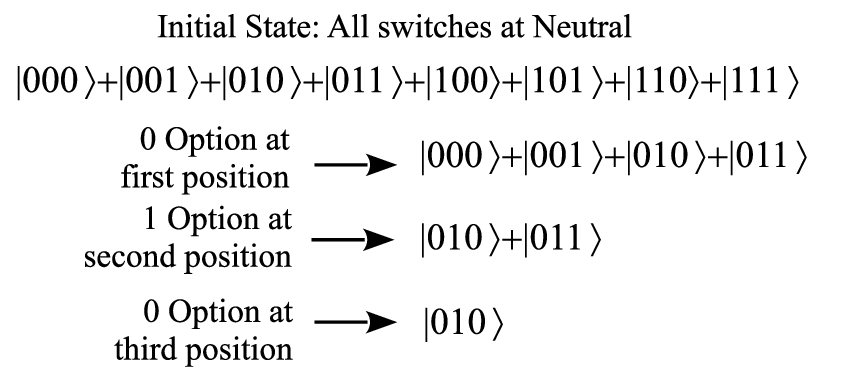} 
\caption{A generative determination of the outcome $|010\rangle$}
\label{fig12}
\end{figure}

The initial experience or first letter in the code sets the first switch to
the $0$ option which reduces the state to $|000\rangle+|001\rangle
+|010\rangle+|011\rangle$ (where the plus signs in the superposition of
these options indicate that the second and third switches are still in
neutral). Then subsequent experience sets the second switch to the $1$
option and the third switch to the $0$ option. Thus, we reach the same
outcome $|010\rangle$ as the final outcome in the two models but by quite
different mechanisms. Note that the generative mechanism `selects' or
determines a specific outcome but that does not make it a `selectionist'
mechanism since it is making distinctions to turn an indefinite
superposition-like state into a more definite state, as opposed to selecting
between already existing variations according to a fitness criterion.

Another way to visually compare a selectionist mechanism with a generative
mechanism is to consider a single-elimination (or knockout) tournament as a ``red in tooth and claw'' selectionist
mechanism versus the implementation of a code for a specific leaf as a generative mechanism as
in Figure 13. The selectionist mechanism starts with $8$ existing teams and
then binary contests whittle down the survivors to a eventual winner. The
generative mechanism starts at the root, which like the superposition $%
\mathbf{0}_{U}$, embodies $8$ possibilities and the sequence of binary-code
switches will eventually distinguish the coded leaf. The fundamental
(reverse-the-arrows) duality of category theory is turn-around-the-trees in this case of Figure
13.

\begin{figure}[h]
\centering
\includegraphics[width=0.9\linewidth]{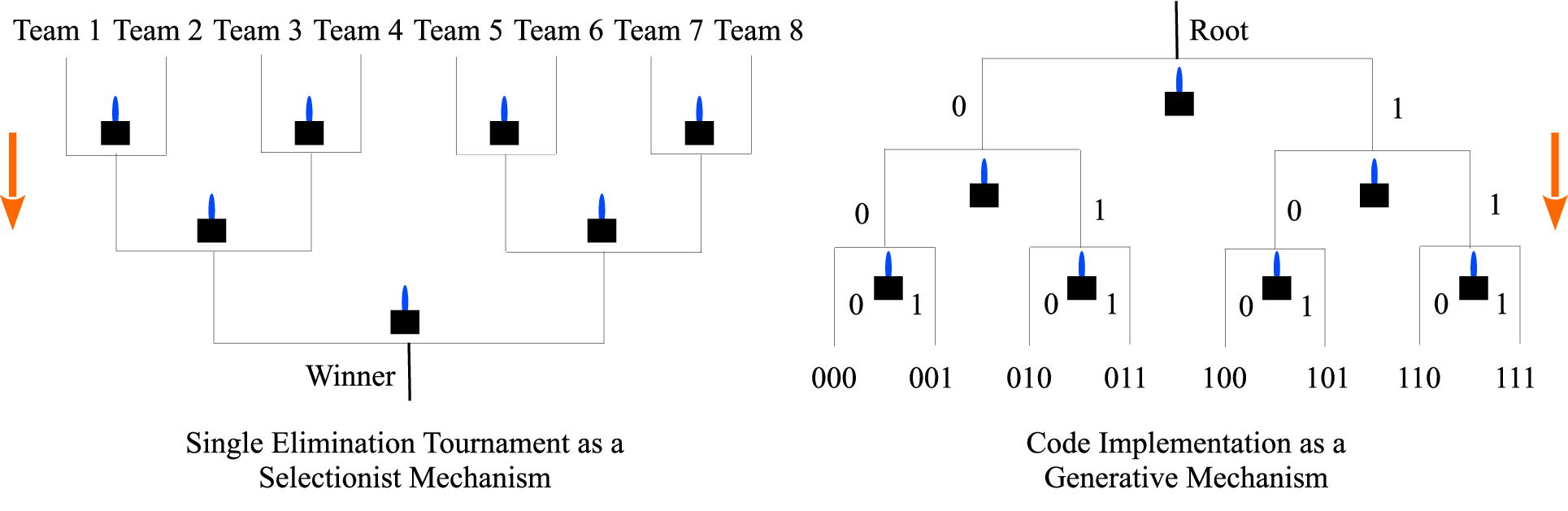} 
\caption{Dual binary selectionist and generative mechanisms}
\label{fig13}
\end{figure}


\section{Discussion and Conclusions}

We have argued that there is a fundamental or foundational duality that runs
through logic, mathematics, probability and information theory, physics, and
even the life sciences. At the logical level, it is the duality between
subsets (or subobjects or `parts') and partitions (or equivalence relations
or quotient objects). At a more granular level, it is the duality between
elements (of a subset) and distinctions (of a partition) or ``Its \& Dits.''
In most cases, there has been a fulsome development of the subset-side of
the duality to the neglect of the partition-side.

\begin{itemize}
\item In logic the developments from the 19th century onwards have started
with the Boolean logic of subsets while partition logic was only developed
in the 21st century \cite{ell:lop2apps}.

\item In mathematics and particularly in category theory, there has been an
even-handed development of both sides of the duality, i.e., subobjects and
quotient objects or limits and colimits, and, in general, the
reverse-the-arrows duality \cite{ell:canonical}.

\item The quantitative versions of subsets and partitions have been
independently developed as probability theory and information theory. But
the information theory was based on Shannon entropy to the neglect of the
more fundamental notion of logical entropy as the quantitative measure of
partitions (\cite{ell:entropy}, \cite{ell:4open}, \cite{ell:nf4it}).

\item In physics, classical physics exemplified the fully-definite view of
reality; an element is definitely in a subset or in its complementary subset
as in the Boolean logic of subsets. Quantum physics developed with the
quantum reality embodying the possibility of objective indefiniteness in
superposition states but the connection with the mathematics of partitions
(or equivalence relations) was only recently understood (\cite{ell:new-theories}, \cite{ell:piqr-book}). Since new jury-rigged interpretations of QM are invented rather often, this approach to understanding QM as the application of a fundamental duality running throughout the exact sciences gives this treatment some cachet above today's ``demolition derby'' of competing interpretations.

\item And in the life sciences, there has long been the emphasis on the
selectionist mechanism which operates on the logic of the existence of
actualized definite alternatives which are then subjected to the ``survival
of the fittest'' criterion. Selectionism was usually juxtaposed to the false
alternative of instructionism or Lamarckism. But the other side of the
duality is the notion of a generative mechanism which we have seen
implemented in a number of biological processes where codes-as-distinctions
guide the process of development of an indefinite state to a definite
outcome (symbolized in the rooted tree diagrams) such as the genetic code in
the DNA-RNA machinery, language acquisition in generative grammar, and
embryonic development from stem cells. \cite{ell:gen-mech}.
\end{itemize}

While the fundamental duality finds its most mathematical formulation as
category theory's reverse-the-arrows duality, that is far too abstract to
elicit the multitude of applications throughout the sciences. The more
specific formulation between subsets and partitions at the logical level,
and the even more granular formulation as the elements-and-distinctions (or
Its \& Dits) duality, brought out many applications--including the `origin'
of the category-theoretic duality in the ur-category of $Sets$. Outside of
category theory, the historical development has been largely on the subset
side of the duality so it was the new developments of the partition side,
starting with partition logic and running through logical information theory, quantum theory, and finally to the biological notion of a generative mechanism, that revealed the wide range of applications of the fundamental duality throughout the mathematical and natural sciences.

\section{Declarations}

The author has received no funding concerning this paper. An institutional
review is not applicable and there are no additional data. DE certifies that
he has no financial conflict of interest (e.g., consultancies, stock
ownership, equity interest, patent/licensing arrangements, etc) in
connection with this article. There are no acknowledgments.


\begin{thebibliography}{99}

\bibitem{law:sfm} Lawvere, F. William and Robert Rosebrugh 2003. \textit{%
Sets for Mathematics}. Cambridge: Cambridge University Press.

\bibitem{ell:lop} Ellerman, David 2010. The Logic of Partitions:
Introduction to the Dual of the Logic of Subsets. \textit{Review of Symbolic
Logic}. 3 (2 June): 287-350.

\bibitem{ell:intropartitions} Ellerman, David. 2014. An Introduction of
Partition Logic. \textit{Logic Journal of the IGPL}. 22 (1): 94--125.

\bibitem{ell:lop2apps} Ellerman, David. 2023. \textit{The Logic of
Partitions: With Two Major Applications. Studies in Logic 101}. London:
College Publications.

\bibitem{ell:nf4it} Ellerman, David. 2021. \textit{New Foundations for
	Information Theory: Logical Entropy and Shannon Entropy}. Cham, Switzerland:
SpringerNature. https://doi.org/10.1007/978-3-030-86552-8.

\bibitem{ell:4open} Ellerman, David. 2022. ``Introduction to Logical Entropy
and Its Relationship to Shannon Entropy.'' \textit{4Open Special Issue:
	Logical Entropy} 5 (1): 1\^{a}\euro ``33..
https://doi.org/10.1051/fopen/2021004.

\bibitem{manfredi:4open} Manfredi, Giovanni, ed. 2022. 
Logical Entropy -- Special Issue. \textit{4Open}, no. 5: E1.
https://doi.org/10.1051/fopen/2022005.

\bibitem{ell:piqr-book} Ellerman, David. 2024.\textit{\ Partitions,
	Objective Indefiniteness, and Quantum Reality: The Objective Indefiniteness
	Interpretation of Quantum Mechanics}. Cham, Switzerland: Springer Nature.
	
\bibitem{ell:gen-mech} Ellerman, David. 2021. \textquotedblleft Generative
	Mechanisms: The Mechanisms That Implement Codes.\textquotedblright\ \textit{%
		ArXiv.org}. https://arxiv.org/abs/2105.03907.

\bibitem{britz:eq-rels} Britz, Thomas, Matteo Mainetti, and Luigi Pezzoli.
2001. \textquotedblleft Some Operations on the Family of Equivalence
Relations.\textquotedblright\ In \textit{Algebraic Combinatorics and
Computer Science: A Tribute to Gian-Carlo Rota}, edited by Henry Crapo and
Domenico Senato, 445--59. Milano: Springer.

\bibitem{birkhoff:lattice-theory3rd} Birkhoff, Garrett. 1973. \textit{%
Lattice Theory 3rd Ed}. Providence RI: American Mathematical Society.

\bibitem{kung:rotaway} Kung, Joseph P. S., Gian-Carlo Rota, and Catherine H.
Yan. 2009. \textit{Combinatorics: The Rota Way}. New York: Cambridge
University Press.

\bibitem{law:conceptual} Lawvere, F. William, and Stephen Schanuel. 1997. 
\textit{Conceptual Mathematics: A First Introduction to Categories}. New
York: Cambridge University Press.

\bibitem{ell:canonical} Ellerman, David. 2022. \textquotedblleft The Logical
Theory of Canonical Maps: The Elements \& Distinctions Analysis of the
Morphisms, Duality, Canonicity, and Universal Constructions in
Set.\textquotedblright\ \textit{ArXiv.org}.\
https://arxiv.org/abs/2104.08583.

\bibitem{rota:fubini} Rota, Gian-Carlo. 2001. \textquotedblleft Twelve
Problems in Probability No One Likes to Bring Up.\textquotedblright\ In 
\textit{Algebraic Combinatorics and Computer Science: A Tribute to
Gian-Carlo Rota}, edited by Henry Crapo and Domenico Senato, 57--93. Milano:
Springer.

\bibitem{campbell:measure} Campbell, L. Lorne. 1965. \textquotedblleft
Entropy as a Measure.\textquotedblright\ \textit{IEEE Trans. on Information
Theory IT-11} (January): 112--14.
	
\bibitem {nei:diversity}Nei, Masatoshi. 1973. Analysis of Gene Diversity in
Subdivided Populations. \textit{Proc. Nat. Acad. Sci. U.S.A.} 70: 3321--3.
	
\bibitem {weir:genetics}Weir, Bruce S. 1996. \textit{Genetic Data Analysis II:
Methods for Discrete Population Genetic Data}. Sunderland MA: Sinauer Associates.	

\bibitem{shannon:comm} Shannon, Claude E. 1948. \textquotedblleft A
Mathematical Theory of Communication.\textquotedblright\ \textit{Bell System
Technical Journal} 27: 379--423; 623--56.

\bibitem{ainworth:form-matter} Ainsworth, Thomas. 2020. \textquotedblleft
Form vs. Matter.\textquotedblright\ In \textit{The Stanford Encyclopedia of
Philosophy (Summer 2020 Edition)}, edited by Edward N. Zalta.
https://plato.stanford.edu/archives/sum2020/entries/form-matter/.

\bibitem{heisenberg:nuclear} Heisenberg, Werner. 1952. \textit{Philosophic
Problems of Nuclear Science}. Greenwich CN: Fawcett Publications.

\bibitem{pagels:time} Pagels, Heinz. 1985. \textit{Perfect Symmetry: The
Search for the Beginning of Time}. New York: Simon and Schuster.

\bibitem{leib-clarke:letters} Ariew, Roger, ed. 2000. \textit{G. W. Leibniz
and Samuel Clarke: Correspondence}. Indianapolis: Hackett.

\bibitem{kant:cpr} Kant, Immanuel. 1998. \textit{Critique of Pure Reason}.
Translated by Paul Guyer and Allen W. Wood. The Cambridge Edition of the
Works of Immanuel Kant. Cambridge UK: Cambridge University Press.

\bibitem{teller:haecceity} Teller, P. 1998. “Quantum Mechanics and Haecceities.” In \textit{Interpreting Bodies: Classical and Quantum Objects in Modern Physics}, edited by E. Castellani, 114–41. Princeton: Princeton University Press.

\bibitem{stachel:hole} Stachel, John. 2014. “The Hole Argument and Some Physical and Philosophical Implications.” \textit{Living Reviews in Relativity} 17:1–66. https://doi.org/doi:10.12942/lrr-2014-1.

\bibitem{bunge:m-and-m} Bunge, Mario. 2010. \textit{Matter and Mind: A
Philosophical Inquiry. Boston Studies in the Philosophy of Science \#287}.
Dordrecht: Spring Publications.

\bibitem{cushing:qft} Cushing, James T. 1988. \textquotedblleft Foundational
Problems in and Methodological Lessons from Quantum Field
Theory.\textquotedblright\ In \textit{Philosophical Foundations of Quantum
Field Theory}, edited by Harvey R. Brown and Rom Harre, 25--39. Oxford:
Clarendon Press.

\bibitem{ell:ftm} Ellerman, David. 2022. \textquotedblleft Follow the Math!:
The Mathematics of Quantum Mechanics as the Mathematics of Set Partitions
Linearized to (Hilbert) Vector Spaces.\textquotedblright\ \textit{%
Foundations of Physics} 52 (5). https://doi.org/10.1007/s10701-022-00608-3.

\bibitem{ell:new-theories} Ellerman, David. 2024. \textquotedblleft A New
Logic, a New Information Measure, and a New Information-Based Approach to
Interpreting Quantum Mechanics.\textquotedblright\ \textit{Entropy. Special
Issue: Information-Theoretic Concepts in Physics} 26 (2).
https://doi.org/10.3390/e26020169.

\bibitem{auletta:qm} Auletta, Gennaro, Mauro Fortunato, and Giorgio Parisi.
2009. \textit{Quantum Mechanics}. Cambridge UK: Cambridge University Press.

\bibitem{luders:meas} L\"{u}ders, Gerhart. 1951. \textquotedblleft\"{U}ber
Die Zustands\"{a}nderung Durch Me\ss proze\ss .\textquotedblright\ \textit{%
Annalen Der Physik} 8 (6): 322--28. Trans. by K. A. Kirkpatrick at:
arXiv:quant-ph/0403007.

\bibitem{furry:luders} Furry, W. H. 1966. \textquotedblleft Some Aspects of
the Quantum Theory of Measurement.\textquotedblright\ In \textit{Lectures in
Theoretical Physics Vol. 8A}, edited by Wesley E. Brittin, 1--64. Boulder
CO: University of Colorado Press.

\bibitem{jaeger:qobjects} Jaeger, Gregg. 2014. \textit{Quantum Objects:
Non-Local Correlation, Causality and Objective Indefiniteness in the Quantum
World}. Heidelberg: Springer.

\bibitem{feynman:1951} Feynman, Richard P. 1951. The Concept of Probability
in Quantum Mechanics.\ In , 533--41. \textit{Second Berkeley Symposium on
Mathematical Statistics and Probability}. University of California Press.

\bibitem{dirac:principles} Dirac, Paul A. M. 1958. \textit{The Principles of
Quantum Mechanics (4th ed.)}. Oxford: Clarendon.

\bibitem{weyl:pmns} Weyl, Hermann. 1949. \textit{Philosophy of Mathematics
and Natural Science}. Princeton NJ: Princeton University Press.

\bibitem{weyl:groups-qm} Weyl, Hermann. 1950. \textit{The Theory of Groups
and Quantum Mechanics}. Translated by H. P. Robertson. New York: Dover
Publications.

\bibitem{chen:group-reps} Chen, J.-Q., J. Ping, and F. Wang. 2002. \textit{%
Group Representation Theory for Physicists (2nd Ed.)}. Singapore: World
Scientific.

\bibitem{dennett:darwin} Dennett, Daniel. 1995. \textit{Darwin's Dangerous
Idea: Evolution and the Meanings of Life}. New York: Touchstone.

\bibitem{cziko:womiracles} Cziko, Gary. 1995. \textit{Without Miracles:
Universal Selection Theory and the Second Darwinian Revolution}. Cambridge:
MIT Press (A Bradford Book).

\bibitem{martin-england:entropy} Martin, Nathaniel, and James England. 1981. 
\textit{Mathematical Theory of Entropy. Encyclopedia of Mathematics and Its
Applications Vol. 12}. Reading MA: Addison-Wesley.

\bibitem{chomp-lasnik:pandp} Chomsky, Noam, and Howard Lasnik. 1993.
\textquotedblleft The Theory of Principles and
Parameters.\textquotedblright\ In \textit{Syntax: An International Handbook
of Contemporary Research}, edited by J. Jacobs, A. von Stechow, W.
Sternefeld, and T. Vennemann, 506--69. Berlin: de Gruyter.

\bibitem{chomsky:minimalist} Chomsky, Noam. 1995. \textit{The Minimalist
Program}. Cambridge: MIT Press.

\bibitem{higginbotham} Higginbotham, James. 1982. \textquotedblleft Noam
Chomsky's Linguistic Theory.\textquotedblright\ \textit{Social Research} 49
(1): 143--57.

\bibitem{chomsky:knowoflang} Chomsky, Noam. 1986. \textit{Knowledge of
Language: Its Origin, Nature, and Use}. New York: Praeger.

\bibitem{roberts:ug} Roberts, Ian. 2019. \textit{Parameter Hierarchies \&
Universal Grammar}. Oxford UK: Oxford University Press.

\bibitem{skinner:behave} Skinner, B. F. 1976. \textit{About Behaviorism}.
New York: Vintage Books.

\bibitem{chomp-mcgil:interview} Chomsky, Noam, and James McGilvray, eds.
2012. \textit{The Science of Language: Interviews with James McGilvray}.
Cambridge UK: Cambridge University Press.

\bibitem{nowak-komarove} Nowak, M. A., and N. L. Komarova. 2001. Towards an
Evolutionary Theory of Language. \textit{Trends in Cognitive Sciences}. 5 (7
July): 288-95.

\bibitem{cepel:mathcells} Cepelewicz, Jordana. 2019. \textquotedblleft The
Math That Tells Cells What They Are.\textquotedblright\ \textit{Quanta
Magazine (blog)}. March 13, 2019.
https://www.quantamagazine.org/the-math-that-tells-cells-what-they-are-20190313/ .

\bibitem{zernicka:dance} Zernicka-Goetz, Magdalena, and Roger Highfield.
2020. \textit{The Dance of Life: The New Science of How a Single Cell
Becomes a Human Being}. New York: Basic Books.

\bibitem{medawar:reith} Medawar, Peter B. 1960. \textit{The Future of Man:
Reith Lectures 1959}. London: Methuen.

\bibitem{jerna:antibodies} Jerne, Niels 1967. Antibodies and learning:
Selection versus instruction. In \textit{The neurosciences: A study program}%
. G. C. Quarton, T. Melnechuk and F. O. Schmitt eds., New York: Rockefeller
University Press: 200-5.

\bibitem{jerne:nat-sel} Jerne, Niels. 1955. The natural selection theory of
antibody formation. \textit{Proc. National Academy of Sciences U.S.A.} 41,
849.

\bibitem{jerne:nobel} Jerne, Niels 1985. The Generative Grammar of the
Immune System. \textit{Science}. 229 (4718): 1057-59.

\bibitem{alberts-et-al:mbio} Alberts, B., D. Bray, J. Lewis, M. Raff, K.
Roberts, and J.D. Watson. 1994. \textit{Molecular Biology of the Cell}. New
York: Garland.

\bibitem{ell:entropy} Ellerman, David. 2018. ``Logical Entropy: Introduction
to Classical and Quantum Logical Information Theory.'' \textit{Entropy} 20
(9): Article ID 679. https://doi.org/10.3390/e20090679.


\end{thebibliography}
\end{document}